\documentclass[reprint,twocolumn,longbibliography,superscriptaddress,
showpacs,preprintnumbers,
notitlepage,amsmath,amssymb,
prx,
]{revtex4-1}

\usepackage{graphicx}
\usepackage{dcolumn}
\usepackage{bm}
\usepackage{bbm}
\usepackage{diagbox}
\usepackage{footnote}
\usepackage{braket}

\usepackage{graphicx}
\usepackage{enumerate}
\usepackage{subfigure}
\usepackage{epstopdf}
\usepackage[colorlinks,
linkcolor=red,
anchorcolor=black,
citecolor=blue]{hyperref}
\usepackage{amssymb}
\usepackage{framed}
\usepackage{tabularx}
\usepackage{booktabs}
\usepackage{float}
\usepackage[table]{xcolor}
\usepackage{array}
\usepackage{tikz}
\hypersetup{colorlinks=true}
\usepackage{color}
\usepackage{wasysym}
\usepackage{mathtools}
\usepackage{comment}
\usetikzlibrary{through,hobby}
\usetikzlibrary{decorations.markings}
\usetikzlibrary{fadings,shadings}
\definecolor{darkblue}{rgb}{0.,0.,0.4}
\definecolor{darkred}{rgb}{0.5,0.,0.}
\definecolor{BlueViolet}{RGB}{138,43,226}
\definecolor{SkyBlue}{RGB}{30,144,255}
\definecolor{DarkGreen}{RGB}{0,150,0}
\definecolor{DarkYellow}{RGB}{0,1,1}
\usetikzlibrary{intersections}
\definecolor{iro100}{cmyk}{1,0,0,0}
\definecolor{iro90}{cmyk}{.9,0,0,0}
\definecolor{iro80}{cmyk}{.8,0,0,0}
\definecolor{iro60}{cmyk}{0,.6,0,0}
\definecolor{iro10}{cmyk}{0,.1,0,0}
\makeatletter
\newsavebox{\@brx}
\newcommand{\llangle}[1][]{\savebox{\@brx}{\(\m@th{#1\langle}\)}%
  \mathopen{\copy\@brx\kern-0.5\wd\@brx\usebox{\@brx}}}
\newcommand{\rrangle}[1][]{\savebox{\@brx}{\(\m@th{#1\rangle}\)}%
  \mathclose{\copy\@brx\kern-0.5\wd\@brx\usebox{\@brx}}}
\makeatother

\newcommand{\1}{\text{\uppercase\expandafter{\romannumeral1}}}
\newcommand{\2}{\text{\uppercase\expandafter{\romannumeral2}}}
\newcommand{\3}{\text{\uppercase\expandafter{\romannumeral3}}}
\newcommand{\4}{\text{\uppercase\expandafter{\romannumeral4}}}
\newcommand{\5}{\text{\uppercase\expandafter{\romannumeral5}}}
\newcommand{\6}{\text{\uppercase\expandafter{\romannumeral6}}}

\def\U{U(1)}
\def\H{\mathcal{H}}
\def\L{\mathcal{L}}
\def\N{\mathcal{N}}
\def\T{\mathcal{T}}
\def\Z{\mathbb{Z}}
\def\E{\mathcal{E}}

\begin{document}

\title{Fractonic Higher-Order Topological Phases in Open Quantum Systems}

\author{Jian-Hao Zhang}
\thanks{These authors contributed equally}
\affiliation{Department of Physics, The Pennsylvania State University, University Park, Pennsylvania 16802, USA}
\author{Ke Ding}
\thanks{These authors contributed equally}
\affiliation{State Key Laboratory of Low-Dimensional Quantum Physics, Department of Physics, Tsinghua University, Beijing 100084, China}
\author{Shuo Yang}
\email{shuoyang@tsinghua.edu.cn}
\affiliation{State Key Laboratory of Low-Dimensional Quantum Physics, Department of Physics, Tsinghua University, Beijing 100084, China}
\affiliation{Frontier Science Center for Quantum Information, Beijing, China}
\affiliation{Hefei National Laboratory, Hefei 230088, China}
\author{Zhen Bi}
\email{zjb5184@psu.edu}
\affiliation{Department of Physics, The Pennsylvania State University, University Park, Pennsylvania 16802, USA}

\begin{abstract}
In this work, we study the generalization of decohered average symmetry-protected topological phases to open quantum systems with a combination of subsystem symmetries and global symmetries. In particular, we provide examples of two types of intrinsic average higher-order topological phases with average subsystem symmetries. A classification scheme for these phases based on generalized anomaly cancellation criteria of average symmetry is also discussed. 
\end{abstract}

\maketitle

\section{Introduction}
The rapid advancement of quantum simulators has sparked interdisciplinary research on the creation and manipulation of entangled quantum states within the noisy intermediate-scale quantum (NISQ) platforms. This development has attracted significant attention from both the condensed matter and quantum information communities \cite{Preskill_2018, Bernien_2017, Ruben_2023}. Symmetry-protected topological (SPT) phases \cite{PhysRevB.80.155131, doi:10.1126/science.1227224, Chen_2014,  PhysRevB.87.155114, Senthil_2015, Vishwanath_2013, Wang_2013, Wang_2014} serve as a class of quantum states with nontrivial quantum entanglements and anomalous boundary states that have great opportunities to be realized in quantum devices, and they provide the resource states for measurement-based quantum computation and preparations for other highly entangled quantum states \cite{Briegel_2001, Raussendorf_2005, Briegel_2009, Aguado_2008, Bolt_2016, Stephen_2017, Raussendorf_2019, Tim_2022, JYLee_2022b, Zhu_2022, Tim_2023a, Guo_2023}. In particular, symmetry-protected topological phases with subsystem symmetries, a novel type of symmetry whose conserved charges are localized in rigid submanifolds of the whole system, have been shown to have practical advantages in realizing measurement-based quantum computations in certain schemes \cite{Devakul_2018, Stephen_2019, Daniel_2020, Roberts_2020}. 

From the perspective of condensed matter physics, subsystem symmetries present a fascinating opportunity to explore new forms of matter characterized by fractonic dynamics of their excitations \cite{Vijay_2016, Nandkishore_2019, Pretko_2020, Pai_2019}. These symmetries also offer a valuable platform for studying strongly interacting topological phases, as the usual single particle hoppings are typically prohibited. Currently, there is active research into the classification and physical properties of subsystem symmetry-protected topological (SSPT) phases \cite{PhysRevB.98.035112, Devakul_2019, PhysRevB.98.235121, PhysRevResearch.2.012059, Williamson_2019, May_Mann_2019, PhysRevB.98.035112, Stephen_2020, May_Mann_2021, PhysRevB.105.245122, 2022arXiv221015596Z}. Recently, a novel class of SSPT phases called fractonic higher-order topological phases has been introduced \cite{PhysRevB.105.245122, 2022arXiv221015596Z}. These topological phases exhibit symmetry-protected gapless modes that manifest only on specific lower-dimensional subspaces of the boundary, while the rest of the boundary remains gapped in a manner consistent with the symmetry. They serve as analogs to higher-order topological phases found in systems with crystalline symmetries \cite{PhysRevLett.106.106802, Hsieh_2012, PhysRevB.92.081304, Tang_2019, Kruthoff_2017, Slager_2012, Bultinck_2019, Laubscher_2019,Thorngren_2018,PhysRevX.7.011020, Huang_2017, PhysRevB.101.100501, PhysRevResearch.4.033081, PhysRevB.106.L020503, Zhang:2022zbp, 3DcSPT}. However, these phases are inherently strongly interacting due to the presence of subsystem symmetries.

Although discussing SSPT phases in the context of the ground state is fascinating, for practical purposes, it is crucial to consider the impact of decoherence and/or dissipation on the quantum entanglement of these topological phases, as systems are inevitably coupled to environments \cite{Fulga_2014, Milsted_2015, Kimchi_2018, de_Groot_2022}. Understanding whether symmetry-protected topological phases remain stable under such conditions is an intriguing and significant question. Specifically, decoherence and/or dissipation can break exact symmetries in closed systems, resulting in an average symmetry in open systems \cite{MaWangASPT, JYLee_2022, Zhang_2022, average_2023}. Consequently, investigating average SPT (ASPT) phases becomes crucial, particularly in NISQ platforms, where quantum dynamics is not solely governed by Hamiltonians. Recent findings indicate that a wide range of SPT persists in mixed-state settings \cite{MaWangASPT}, and notably, there are numerous nontrivial SPTs whose existence requires the assistance of quantum decoherence, referred to as intrinsic ASPT phases \cite{average_2023}. 


In this paper, we examine the higher-order subsystem symmetry-protected topological (SSPT) phases in strongly correlated open systems subjected to quantum decoherence. For concreteness, we will mostly consider $(3+1)d$ systems with two-foliated subsystem symmetries. After some introductory remarks in this section, we will introduce examples of higher-order subsystem average SPT (SASPT) phases, and in particular, we will discuss two types of intrinsic SASPT phases with examples. Then we will present a general scheme for the classification of the higher-order SASPT phases in Sec. \ref{Sec: classification} using the idea of anomaly cancellation for average symmetries. In addition to exploring the systematic classification and construction scheme of SASPT states, we also study examples of dynamical phase transitions of the hinge modes due to the effects of decoherence or dissipation in Sec. \ref{Sec: dynamics}. We conclude and discuss some future outlooks in Sec. \ref{Sec: Summary}. In the appendix, some detailed peripheral discussion about SASPT is provided. 

\begin{figure}
\centering\includegraphics[width = 0.5\linewidth]{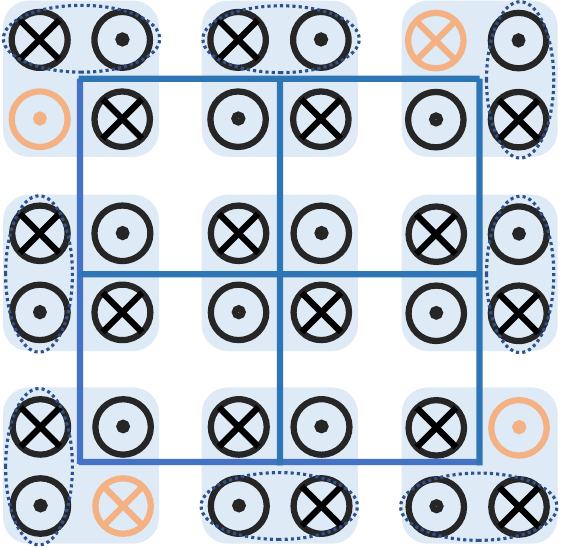}
\caption{Coupled-wire construction of two-foliated subsystem symmetries. Each ``$\bigotimes$" and ``$\bigodot$" represent a pair of modes that carry opposite anomalies of certain symmetries labeled by $[\nu]$ and $[-\nu]$. The orange circles represent the second-order gapless hinge modes.}
\label{fig1}
\end{figure}

\subsection{Higher-order SSPT in clean systems}
For higher-order SSPT in clean systems \cite{Devakul_2019}, there has been a systematic way to classify and construct these phases in all dimensions for bosonic systems \cite{2022arXiv221015596Z} which is based on the idea of anomaly cancellation. We will take a $(3+1)d$ system with two-foliated subsystem symmetry as an example to illustrate the idea. Suppose that we have a system that has a finite extension in the $x$ and $y$ directions while being infinite in the remaining direction. We assume subsystem symmetries along every $xz$ and $yz$ plane of the system. By definition, the boundary of this system is trivially gapped except for the four hinges. These hinge modes are protected by the subsystem symmetries (and possibly some global symmetry as well). In other words, each individual hinge mode carries anomalies of these symmetries. However, when we view this whole system as a quasi-one-dimensional system, with the subsystem symmetries now becoming on-site symmetry groups, clearly, as a physical one-dimensional system, the symmetry actions must be free of any 't Hooft anomalies. This is the consistency condition that we need to impose to classify the SSPT phases. 
For higher-order SSPT in clean systems \cite{Devakul_2019}, there has been a systematic way to classify and construct these phases in all dimensions for bosonic systems \cite{2022arXiv221015596Z} which is based on the idea of anomaly cancellation. We will take a $(3+1)d$ system with two-foliated subsystem symmetry as an example to illustrate the idea. Suppose that we have a system that has a finite extension in the $x$ and $y$ directions while being infinite in the remaining direction. We assume subsystem symmetries along every $xz$ and $yz$ plane of the system. By definition, the boundary of this system is trivially gapped except for the four hinges. These hinge modes are protected by the subsystem symmetries (and possibly some global symmetry as well). In other words, each individual hinge mode carries anomalies of these symmetries. However, when we view this whole system as a quasi-one-dimensional system, with the subsystem symmetries now becoming on-site symmetry groups, clearly, as a physical one-dimensional system, the symmetry actions must be free of any 't Hooft anomalies. This is the consistency condition that we need to impose to classify the SSPT phases. 

Note that this picture also automatically gives a ``coupled-wire" construction for the SPT phase. In each unit cell of the coupled-wire construction, we put the four anomalous hinge modes together, which we refer to as four building blocks throughout the paper. Within a unit cell, the four building blocks together are anomaly free and can be realized in purely one dimension. Then we consider some interacting Hamiltonian between the unit cells that preserve the subsystem symmetries. The anomaly cancellation condition is equivalent to the statement that the bulk of the system can be symmetrically gapped out by turning on symmetric interactions between the neighboring four unit cells. The hinge of the system, however, is left gapless because there are no modes to pair up with the assumptions of locality and symmetry constraints. A schematic illustration for the above physical picture is shown in Fig. \ref{fig1}.

More formally, the bosonic anomaly for the group $G$ in $(d+1)$ space-time dimensions is characterized by a cocycle in $d+2$ dimension, namely $[\nu]\in \H^{d+2}[G, \U]$ where $\nu$ is a representative group cocycle. Therefore, each anomalous hinge mode for two-foliated higher-order SSPT in $d$-spatial dimensions is labeled by a nontrivial cocycle in $\H^{d}[G_s, \U]$ where $G_s$ is the subsystem symmetry group. Note that for each individual hinge, only a limited set of subsystem symmetry is involved. The anomaly-free condition is when we take the cocycles on the four hinges together and consider the full subsystem symmetry groups the system should carry no anomaly. It turns out this condition is equivalent to saying that the image of the following map between $\H^{d}[G_s, \U]$ and $\H^{d}[G_s\times G_s, \U]$,
\begin{align}
f_2(\nu) (\{g,g'\})=\frac{\nu(\{g\})\nu(\{g'\})}{\nu(\{gg'\})},
\label{two-foliated anomaly free}
\end{align}
must be the trivial class in $\H^{d}[G_s\times G_s, \U]$. 
We note that the above program based on anomaly cancellation is also applicable to fermionic systems. The anomalies of fermion SPT phases are described by generalized group cohomology \cite{Wang_2018, Wang_2020}, and we only need to substitute the cocycle $\nu$ in Eq. \eqref{two-foliated anomaly free} by generalized cocycle describing the fermionic SPT anomaly for fermionic systems.

These conditions can be easily generalized to $n$-foliated structures in general dimensions and to incorporate additional global symmetries. For interested readers, we refer to more details in Ref. \cite{2022arXiv221015596Z}. We will generalize these ideas into the classification of SASPT phases with decoherence in Sec. \ref{Sec: classification}.

\subsection{Coupled-wire construction}
\label{Sec: closed coupled-wire}
As mentioned above, the coupled-wire construction is a natural physical picture for SSPT in $(3+1)d$ with two-foliated subsystem symmetries. Here, we will rephrase the anomaly-free condition for constructing SSPT in this setting.  As illustrated in Fig. \ref{fig1}, we have four building blocks per unit cell, and each block is an anomalous $(1+1)d$ edge state of certain $(2+1)d$ SPT state classified by $\nu\in H^3[G_s, U(1)]$. The arrangement of the four anomalous modes in one unit cell is $\nu,\nu^{-1},\nu,\nu^{-1}$. The total anomaly of four building blocks in each unit cell is automatically canceled, therefore, the unit cell in principle can be realized in $(1+1)$-dimension. The anomaly cancellation condition described in the preceding section means physically one can get a symmetric gapped bulk state by turning on symmetric interactions between four neighboring unit cells. This condition can be checked using techniques in the framework of $(1+1)d$ multi-channel Luttinger liquids \cite{Kane_2002, Teo_2014}

Considering the four modes in the intersection of four neighboring unit cells, i.e. the modes within a blue plaquette in Fig. \ref{fig1}, the free part of Lagrangian has the following form
\begin{align}
\mathcal{L}_0 = \partial_t{\Phi}^T\frac{{K}}{4\pi}\partial_x{\Phi}+\partial_x{\Phi}^T\frac{{V}}{4\pi}\partial_x{\Phi},
\label{Luttinger}
\end{align}
where $K$ is the $K$-matrix dubbing the topological term, while non-universal $V$ matrix dubbing the dynamical term. The goal to obtain a symmetric gapped bulk becomes to find a complete set of Higgs terms, 
\begin{align}
\L_{\rm Higgs}=\sum_k\cos{(l_{k}^T {K}{\Phi})},
\end{align}
as backscattering of the boson fields in this set of modes that respects all relevant subsystem symmetries. 

For a complete set of Higgs terms, we first require all terms in the set to be mutually commuting. Thus $\{l_k\}$ should satisfy the ``null-vector" condition \cite{PhysRevLett.74.2090}
\begin{equation}\label{null-vector}
l_{i}^T K l_{j}=0, \quad \forall i,j.
\end{equation}
Also, we demand that the number of Higgs terms be the same as the number of helical modes in Eq. \ref{Luttinger} in order to gap all the gapless modes. In the end, we also need to make sure that there is no spontaneous symmetry breaking in the strong coupling limit of these Higgs terms by requiring the minors of the matrix formed by the $l$ vectors to be 1 \cite{PhysRevB.104.075151}. If we can find such a set of Higgs terms, then the bulk of the system can be fully gapped out and the anomaly-free condition is checked. One can easily spot the remaining gapless mode on the hinge of the system, which is exactly the second-order gapless hinge mode of the SSPT state.

\subsection{Topological phases with average symmetries}

For an open quantum system, we are generally concerned with mixed density matrices. The generalization of short-range-entangled (SRE) pure states to mixed states are density matrices that can be prepared from a pure product state using a finite-depth local quantum channel, namely
\begin{align}
\rho=\mathcal{E}(\ket{0}\bra{0}),
\end{align}
where the quantum channel $\E$ can be formulated as
\begin{align}
\E[\rho]=\sum\limits_jK_j\rho K_j^\dag,~\sum\limits_jK_j^\dag K_j=\mathbbm{1},
\end{align}
where $K_j$'s are the local symmetric Kraus operators. An SRE mixed state generically has short-range correlations for all local operators. We would like to discuss symmetry-protect topological phases in such SRE density matrices. 

For the density matrix, two types of symmetries can arise, namely the exact and average symmetries \cite{de_Groot_2022, Fulga_2014, MaWangASPT, average_2023}. We label the exact symmetry group by $K$ and the average symmetry group by $G$. 
The exact symmetry $K$ is defined such that for a symmetry operator $U_k$ ($k\in K$), the density matrix $\rho$ is invariant by acting $U_k$ individually on the left or right, say $U_k\rho= e^{i\alpha}\rho$ and $\rho U_k^\dag=e^{-i\alpha}\rho$. The average symmetry $G$ is defined that for a symmetry operator $U_g$ ($g\in G$), the density matrix $\rho$ is generally not invariant when acting $U_g$ on the left or right individually, but invariant when acting $U_g$ on the left and right simultaneously, say $U_g\rho U_g^\dag=\rho$. Generically, the total symmetry group $\tilde{G}$ is an extension of $G$ symmetry by $K$, which can be characterized by certain short exact sequence
\begin{equation}
     1\rightarrow K\rightarrow\tilde{G}\rightarrow G\rightarrow1.
\end{equation}
When encountering a subsystem symmetry, we will add a subscript $s$ to denote it. 

The concept of average symmetry-protected topological (ASPT) phases is based on the equivalence classes of density matrices under symmetric finite-depth local quantum channels. Specifically, two ASPT density matrices, $\rho_1$ and $\rho_2$, are considered equivalent if there exist symmetric finite-depth local quantum channels, $\mathcal{E}_{12}$ and $\mathcal{E}_{21}$, such that $\mathcal{E}_{12}(\rho_1)=\rho_2$ and $\mathcal{E}_{21}(\rho_2)=\rho_1$. The classification of ASPT states is achieved through the use of generalized group cohomology theory, which provides explicit classifications based on decorated domain wall constructions \cite{Chen_2014, Wang_2021}. Roughly speaking, a nontrivial ASPT density matrix can be constructed as a classical collection of wavefunctions with $G$-symmetry defects decorated with exact $K$-symmetry SPTs. Different decoration patterns give different ASPT phases, as one cannot change the decorated $K$-symmetry SPT without using a deep quantum channel. 

Noticeably, a new class of topological phases that only exists in mixed states, dubbed intrinsic ASPT, is discovered in Ref. \cite{average_2023}. The existence of these new phases is due to the modified consistency relation of the generalized cohomology theory. The basic idea is that the Berry phase consistency condition in the cohomology theory for pure-state SPT is no longer required in a mixed state as the Berry phase is not well defined. Thus, there can be decorated domain wall configurations that make no sense in an SRE quantum wavefunction but can exist in a mixed density matrix. General classification of ASPT phases with global exact and average symmetry are flashed out in Ref. \cite{average_2023}. We describe some detailed mathematical structures in the Appendix. \ref{APP: SS}. In the following, we will generalize the idea of ASPT to systems with subsystem symmetries and possibly some global symmetry as well. 

While the quantum channels are the most general definition of dynamics in open quantum systems, a large class of open quantum system dynamics can be described by the Markovian quantum master equation\cite{Sieberer_2016}, which assumes the following form
\begin{align}
\partial_t\rho=\L\rho=-i[H,\rho]+\sum\limits_\alpha\gamma_\alpha\left(L_\alpha^\dag\rho L_\alpha-\frac{1}{2}\left\{L_\alpha^\dag L_\alpha, \rho\right\}\right).
\label{master}
\end{align}
Here the operator $\L$ is called Liouville super-operator and acts on the density matrix $\rho$ from both sides. The quantum jump operators $L_\alpha$ describe the coupling between systems and environment (bath). The non-negative number $\gamma_\alpha$ depicts the intensity of quantum jumps. In this formalism, the exact symmetry is the symmetry that is preserved by each individual quantum jump operator, while the average symmetry is essentially a statistical symmetry of the ensemble of quantum trajectories.  

The Lindbladian evolution constitutes a versatile framework capable of encompassing diverse non-equilibrium dynamics. Notably, the Lindbladian operator, denoted as $L_\alpha$, can be adeptly applied to model scenarios in which the quantum system undergoes (weak) measurements by its surrounding environment either instantaneously or within finite time intervals, as expressed by the expression $L_\alpha(t)\sim\delta(t-t_0)$. Within the context of our research, we employ the terminology measurement or decoherence to characterize this particular manifestation of quantum dynamics. Moreover, the Lindbladian evolution framework finds applicability in characterizing systems engaged in continuous interactions with their external environments, an aspect we identify as dissipation dynamics in our work. Our analysis of the dynamic phase transition caused by either decoherence or dissipation relies on this versatile framework. We will delve into this topic further in Section \ref{Sec: dynamics} of our paper.

\section{subsystem average SPT Phases}
\label{Sec: intrinsic}
In this section, we present the basic idea of SASPT phases through a coupled wire construction, similar to what was done in the context of $(3+1)d$ clean systems with two-foliated subsystem symmetry \cite{2022arXiv221015596Z}. In parallel to the construction of SSPT in clean systems, we arrange four ``average" anomalous $(1+1)d$ modes within each unit cell as depicted in Fig. \ref{fig1}, and aim to achieve a short-range entangled (SRE) bulk state through symmetric interactions and decoherence. The average anomalous $(1+1)d$ modes can be viewed as the boundary of certain $(2+1)d$ ASPT states and may or may not have a purification into an SRE state. Analogous to the clean systems, we will present the average anomaly cancellation condition for SASPT phases, which is equivalent to the emergence of an SRE bulk state through symmetric interactions and decoherence between different unit cells.

\subsection{SASPTs with clean limits}
\label{Sec: coupled-wire}
First, we illustrate an example of SASPT with a clean limit. In this type of SASPT, since the bulk is already gapped in the clean limit, all we need to show essentially is that the hinge mode of the clean SSPT is stable against decoherence that turns part of the symmetry from exact to average. Consider a $(3+1)d$ bosonic system possessing a two-foliated subsystem symmetry $\mathbb{Z}_2$ and a global time-reversal symmetry $\mathbb{Z}_2^T$. We will first show that, in the clean limit, there is a nontrivial SSPT via wire construction. Then we will argue that, with decoherence that breaks the subsystem symmetry to average while keeping the exact $\mathbb{Z}_2^T$ symmetry, the nontrivial hinge modes cannot be turned into an SRE mixed state. Therefore, this SSPT phase is stable in open systems. 

The wire construction is shown in Fig. \ref{fig1}. Each blue circle represents an anomalous theory carrying the 't Hooft anomaly of a $(2+1)d$ $\Z_2\times\Z_2^T$ SPT state in clean systems. To show that an SSPT exists, the first objective is to introduce symmetric interactions between a different unit cell such that the bulk of the system can be gapped out without breaking any symmetry -- which is the essence of the anomaly cancellation condition. Then one needs to check if there are any nontrivial hinge modes. Finally, if we allow any local quantum channels that break the subsystem $\Z_2$ symmetry to an average subsystem $\Z_2$ symmetry, we will show that the hinge modes remain nontrivial, and hence the system is a SASPT state.

To consider the gapping problem in the bulk, we need to write down the Lagrangian for each plaquette in Fig. \ref{fig1}. This can be conveniently presented as eight-component bosonic fields $\Phi=(\Phi_{1},\cdots,\Phi_{4})$, where $\Phi_{i}=(\phi_{i1},\phi_{i2})$ denotes the bosonic degrees of freedom of the quantum wires in each plaquette. The kinetic part of the Luttinger liquid reads \cite{PhysRevB.86.125119, PhysRevB.104.075151}
\begin{align}
\mathcal{L}_0 = \partial_t{\Phi}^T\frac{{K}}{4\pi}\partial_x{\Phi}+\partial_x{\Phi}^T\frac{{V}}{4\pi}\partial_x{\Phi},
\label{double Luttinger}
\end{align}
where ${K}=(\sigma^x)^{\oplus 4}$ is the K-matrix. Each block of $\sigma^x$ is supposed to be the edge theory of a $(2+1)d$ $\mathbb{Z}_2\times\mathbb{Z}_2^T$ bosonic SPT, which gives the following symmetry transformations
\begin{align}
g:\Phi\rightarrow (-1)^{s_1(g)}W\Phi+\delta\Phi.
\end{align}
Here $s_1(g)$ characterizes if $g$ is anti-unitary, namely
\begin{align}
s_1(g)=\left\{
\begin{aligned}
&0,~~\mathrm{if}~g~\mathrm{is~unitary}\\
&1,~~\mathrm{if}~g~\text{is~anti-unitary}
\end{aligned}
\right.
.
\end{align}
In the plaquette, there are four independent $\mathbb{Z}_2$ subsystem symmetries. Their symmetry transformation rules are given by the following
\begin{align}\begin{aligned}
    W^{\mathbb{Z}_2} &= (\sigma^0)^{\oplus4} \\
    \delta\Phi^{\mathbb{Z}_2^{(1)}} &= \pi(1,0,1,0,0,0,0,0)^T \\
    \delta\Phi^{\mathbb{Z}_2^{(2)}} &= \pi(0,0,0,0,1,0,1,0)^T \\
    \delta\Phi^{\mathbb{Z}_2^{(3)}} &= \pi(1,0,0,0,0,0,1,0)^T \\
    \delta\Phi^{\mathbb{Z}_2^{(4)}} &= \pi(0,0,1,0,1,0,0,0)^T
\end{aligned}.
\end{align}
For the global time-reversal symmetry $\mathcal{T}\in \mathbb{Z}_2^T$
\begin{align}\begin{aligned}
W^{\mathcal{T}} &= (-\sigma^z)^{\oplus4} \\\delta\Phi^{\mathcal{T}} &= \pi(0,1,0,1,0,1,0,1)^T 
\end{aligned}.
\end{align}
To get an SSPT state, we should fully gap each plaquette in the bulk with symmetric interactions. To that end, we need to include four linearly independent symmetric Higgs terms,
\begin{align}
\mathcal{L}_{\rm Higgs}=\sum_{k=1}^4\lambda_k\cos{(l_{k}^T K\Phi)},
\end{align}
that satisfying the null-vector conditions (\ref{null-vector}). One can easily check that the following vectors satisfy all these conditions
\begin{align}
\begin{aligned}
    l_1 &= (1,0,1,0,1,0,1,0)^T \\
    l_2 &= (0,1,0,-1,0,-1,0,1)^T \\
    l_3 &= (1,0,0,-1,0,1,-1,0)^T \\
    l_4 &= (0,1,-1,0,-1,0,0,1)^T
\end{aligned}.
\end{align}
Moving on, we will inspect the properties of the boundary of the system. On a smooth boundary, there exist four dangling bosonic modes. It is obvious that one can introduce on-site mass terms that allow us to fully gap out the boundary site. This will not be true for a corner site, i.e., there is a dangling hinge mode at each corner. By design, each hinge mode is described by a Lagrangian that is precisely the anomalous boundary of the $(2+1)d$ SPT with $\mathbb{Z}_2\times\mathbb{Z}_2^T$ symmetry. Hence, the whole system comprises an SSPT state. 

Finally, we consider the effect of decoherence, which breaks the $\mathbb{Z}_2$ subsystem symmetry down to an average symmetry, on the hinge modes. Since the hinge mode can be viewed as the edge of $(2+1)d$ SPT with $\mathbb{Z}_2\times\mathbb{Z}_2^T$ symmetry, in hindsight the question is equivalent to asking if 2 + 1d SPT is stable after breaking the exact $\mathbb{Z}_2$ symmetry to an average symmetry. The $(2+1)d$ SPT used in the construction belongs to the nontrivial decoration class where a $\mathbb{Z}_2$ domain wall is decorated with a $(1+1)d$ SPT with $\mathbb{Z}_2^T$ symmetry. According to the general classification derived in Ref. \cite{average_2023}, this SPT is still a nontrivial ASPT when $\mathbb{Z}_2$ symmetry is average and $\mathbb{Z}_2^T$ is kept exact. Therefore, its boundary, carrying an average anomaly, cannot be turned into an SRE mixed state. 

While the argument above is generally valid, one can see this more explicitly with the following microscopic example. In clean systems, the hinge mode would be a $(1+1)d$ Luttinger liquid in the form of Eq. \eqref{double Luttinger}, with the $K$-matrix $K=\sigma^x$ and $\Z_2\times\Z_2^T$ symmetry properties as
\begin{align}
\begin{gathered}
W^{\Z_2}=\mathbbm{1}_{2\times2},~\delta\phi^{\Z_2}=\pi(1,0)^T\\
W^{\T}=-\sigma^z,~\delta\phi^{\T}=\pi(0,1)^T
\end{gathered}.
\end{align}
Then we consider a local quantum channel $\N(x)$ of decoherence that breaks $\Z_2$ to an average symmetry, namely
\begin{align}
\N(x)[\rho_{\mathrm{hinge}}]=(1-p)\rho_{\mathrm{hinge}}+p K(x)\rho_{\mathrm{hinge}}K(x)^\dag,
\end{align}
where $K(x)\sim\cos\phi_1(x)$. It is the lowest order Kraus operator in terms of the $\phi$ fields that breaks $\mathbb{Z}_2$ down to average while keeping the exact $\T$ symmetry. The quantum channel $\N$ has no effect on the correlation function of $\cos\phi_1$ operators, which will remain power-law correlated in the decohered systems. Therefore, after applying the quantum channel $\N$ to break $\Z_2$ to an average symmetry, we obtain a $(1+1)d$ power-law correlated mixed hinge state, which is consistent with the conclusion that the system is a nontrivial SASPT.

\subsection{Intrinsic SASPT phases}
In the case of on-site symmetry, it has been proposed that a significant class of ASPT is only well-defined in mixed states, lacking any counterpart in clean (closed) systems \cite{average_2023}. In this subsection, we demonstrate that a large class of SASPT phases also does not have clean limits and can only exist in mixed ensembles. We refer to these SASPT phases as intrinsic SASPT phases. Based on the wire construction picture, we find two types of intrinsic SASPT phases depending on the properties of the building blocks. In the following, we primarily discuss the cases of three-dimensional systems with two-foliated subsystem symmetries for clarity, although general cases are easy to construct. 

\begin{enumerate}[1.]
\item Type-\1 intrinsic SASPT refers to the following situation. We consider coupled-wire models where each unit cell is composed of four building blocks consisting of anomalous modes which can be viewed as the $(1+1)d$ boundaries of certain $(2+1)d$ clean SPTs. The modes are arranged such that within each unit cell, they can be symmetrically gapped out in the clean limit, meaning the system admits a trivial ``atomic" insulating phase in the clean limit. However, we want to consider the situation where it is not possible to gap out the modes from four neighboring unit cells in the clean limit using only symmetric interactions, namely a clean SSPT does not exist. In such a situation, if symmetric local decoherence can lead to an SRE bulk mixed state and leave the hinge of the system nontrivial, we will call such a system a Type-\1 intrinsic SASPT.

\item Type-\2 intrinsic SASPT is different from Type-\1 in that the building blocks in each unit cell do not have a clean limit. In particular, the building blocks are density matrices corresponding to the $(1+1)d$  anomalous boundary of a $(2+1)d$ intrinsic ASPT state \cite{average_2023}. The four building blocks are arranged such that the average anomaly cancels within the unit cell, meaning that the unit cell can be an SRE mixed state with average symmetries. Under such conditions, if symmetric interactions and decoherence can turn the four building blocks from four neighboring unit cells to an SRE mixed state and leave a nontrivial hinge, then we refer to such systems as Type-\2 intrinsic SASPTs.

\end{enumerate}
We note in the type-\1 intrinsic SASPT, the system still admits a clean trivial insulator state, but a nontrivial SPT state exists only in a mixed ensemble. However, in a type-\2 system, the existence of a trivial insulator already requires the system to be an open quantum system. 

\subsubsection{Type-\1 intrinsic SASPT from decoherence-assisted anomaly cancellation}
\label{sec: type-1 intrinsic}
In this section, we focus on intrinsic SASPT states whose anomaly cancellation conditions can only be fulfilled in open quantum systems.

Consider the coupled-wire model discussed in Sec. \ref{Sec: coupled-wire}. Each unit cell consists of four building blocks (depicted in Fig. \ref{fig1}). Each building block serves as an anomalous edge state of a clean SPT state. In the discussion of clean SSPT, one requires an anomaly cancellation condition in Eq. \eqref{two-foliated anomaly free} which can guarantee a symmetric gapped bulk by inter-unit-cell interactions. In the case of type-\1 intrinsic SASPT, however, the anomaly cancellation condition (\ref{two-foliated anomaly free}) does not hold in clean systems. Specifically, the type-\1 intrinsic SASPT states correspond to elements in $H^p[G_s, h^q(K_s)]$ whose image under the $f_2$ map in Eq. (\ref{two-foliated average anomaly cancellation}) is an element in $H^{p+q}[G_s^2, U(1)]$. However, when breaking the subsystem symmetries down to average symmetry by local decoherence channel, such anomaly vanishes. Hence, an SRE bulk is possible in open quantum systems. 

Now we present an example of a three-dimensional system with a two-foliated subsystem symmetry $G_s=\mathbb{Z}_2$ and global fermion parity conservation $G_g=\mathbb{Z}_2^f$. Each building block in the coupled-wire model (Fig. \ref{fig1}) corresponds to the edge theory of two copies of $p\pm ip$ superconductor (or the edge of $\mathbb{Z}_8$-classified fermionic Levin-Gu state in $(2+1)d$ with topological index $\nu=2$). The building block can be described by a Luttinger liquid with a $K$-matrix $K=\sigma^z$ and $\mathbb{Z}_2$ and the $\mathbb{Z}_2^f$ symmetries actions as the following
\begin{align}
\begin{aligned}
\Z_2:~&W^{\Z_2}=\mathbbm{1}_{2\times2},~\delta\phi^{\Z_2}=\pi(0,1)^T\\
\Z_2^f:~&W^{\Z_2^f}=\mathbbm{1}_{2\times2},~\delta\phi^{\Z_2}=\pi(1,1)^T
\end{aligned}.
\label{block symmetry}
\end{align}
In each unit cell, there are four building blocks with an alternating anomaly pattern such that the total anomaly is trivial. Therefore, each unit cell can emerge in a clean system as an SRE pure state.

Now in order to get a nontrivial state, we want to obtain an SRE bulk by symmetric interaction and/or decoherence between the building blocks within one inter-unit-cell plaquette. The Luttinger liquid theory describing the modes inside the inter-unit-cell plaquette in Fig. \ref{fig1} has the $K$-matrix $K=(\sigma^z)^{\oplus4}$. And there are four $\Z_2$ subsystem symmetries acting on these modes with the following nontrivial actions
\begin{align}
\begin{aligned}
&\Z_2^1:~\begin{aligned}
&W^{g_1}=\mathbbm{1}_{8\times8}\\
&\delta\phi^{g_1}=\pi(1,0,0,1,0,0,0,0)^T
\end{aligned}\\
&\Z_2^2:~\begin{aligned}
&W^{g_2}=\mathbbm{1}_{8\times8}\\
&\delta\phi^{g_2}=\pi(0,0,0,0,0,1,1,0)^T
\end{aligned}\\
&\Z_2^3:~\begin{aligned}
&W^{g_3}=\mathbbm{1}_{8\times8}\\
&\delta\phi^{g_3}=\pi(1,0,0,0,0,1,0,0)^T
\end{aligned}\\
&\Z_2^4:~\begin{aligned}
&W^{g_4}=\mathbbm{1}_{8\times8}\\
&\delta\phi^{g_4}=\pi(0,0,0,1,0,0,1,0)^T
\end{aligned}
\end{aligned}.
\label{Z2 subsystem}
\end{align}
We also demand a global fermion parity conservation
\begin{align}
\Z_2^f:~\begin{aligned}
&W^{\Z_2^f}=\mathbbm{1}_{8\times8}\\
&\delta\phi^{\Z_2^f}=\pi(1,1,1,1,1,1,1,1)^T
\end{aligned}.
\label{global fermion parity}
\end{align}

We can first check that these modes cannot be gapped out by symmetric interaction in the clean limit by calculating the anomaly of these symmetries. An anomaly indicator to detect the $\Z_2$ symmetry anomaly is demonstrated in Ref. \cite{Heinrich_2018}: for a Luttinger liquid with $\Z_2$ symmetries (unitary or anti-unitary), one can always put the $K$-matrix and symmetry actions into the following canonical forms
\begin{align}
K=\left(
\begin{array}{cccc}
A & 0 & B & -B\\
0 & C & D & D\\
B^T & D^T & E & F\\
-B^T & D^T & F^T & E
\end{array}
\right),
\end{align}
\begin{align}
W=\left(
\begin{array}{cccc}
-\mathbbm{1}_{n_--m} & 0 & 0 & 0\\
0 & \mathbbm{1}_{n_+-m} & 0 & 0\\
0 & 0 & 0 & \mathbbm{1}_m\\
0 & 0 & \mathbbm{1}_m & 0
\end{array}
\right),
~\delta\phi=\left(
\begin{array}{cccc}
0\\
\chi_2\\
0\\
0
\end{array}
\right),
\end{align}
where $\mathbbm{1}_{m}$ is an $m\times m$ identity matrix, $n_-$, $m$, and $n_+$ are non-negative integers satisfying $n_++n_-=N$ and $m\leq n_\pm$. An auxiliary vector $\chi_+$ is further defined as
\begin{align}
\chi_+=\left(
\begin{array}{cccc}
0\\
\chi_2+2a\\
\mathrm{diag}(E+F)/2+b\\
\mathrm{diag}(E+F)/2+b
\end{array}
\right),~\forall a, b\in\Z_2.
\end{align}
The anomaly indicator $\nu$ is defined by
\begin{align}
\nu\equiv\frac{1}{2}\chi_+^TK^{-1}\chi_++\frac{1}{4}\mathrm{sig}(K(1-W))~(\mathrm{mod}~2),
\label{anomaly indicator}
\end{align}
where ``sig'' denotes the signature of the matrix. The anomaly-free criterion of $(K, W, \delta\phi)$ based on the anomaly indicator $\nu$ is
\begin{align}
\nu=0~(\mathrm{mod}~2).
\end{align}

For $\Z_2$ subsystem symmetries in Eq. \eqref{Z2 subsystem}, one can show the anomaly indicator of the operations $g_1g_3$, $g_2g_4$, $g_1g_4$, and $g_2g_3$ are non-vanishing, as
\begin{align}
\begin{gathered}
\nu_{g_1g_3}=\nu_{g_2g_4}=1(\mathrm{mod}~2)\\
\nu_{g_1g_4}=\nu_{g_2g_3}=1(\mathrm{mod}~2)
\end{gathered}~.
\label{Z2anomaly}
\end{align}
This implies that the modes in each plaquette exhibit subsystem symmetry anomaly, indicating that the coupled-wire construction described above is obstructed to have a gapped bulk in the clean limit by symmetric interaction due to the failure of the anomaly cancellation condition. Equation \eqref{Z2anomaly} also indicates that the anomaly is bosonic \cite{Heinrich_2018}. In other words, by introducing suitable symmetric interaction to each plaquette, the eight-component Luttinger liquid can be transformed into a 2-component boson field with a $K$-matrix $K=\sigma^x$ and the $\mathbb{Z}_2$ symmetry transformation $W^{\mathbb{Z}_2}=\mathbbm{1}_{2\times2}$ and $\delta\phi^{\mathbb{Z}_2}=\pi(1,1)$, which is precisely the boundary of the $(2+1)d$ bosonic $\mathbb{Z}_2$ SPT, i.e., the Levin-Gu state \cite{Levin-Gu}.

Next, we consider the system subject to decoherence, which breaks down the $\mathbb{Z}_2$ subsystem symmetry to an average $\mathbb{Z}_2$ subsystem symmetry. In this situation, the anomalous bosonic modes in each plaquette can actually be turned into an SRE mixed state, because the anomaly becomes trivial when the $\mathbb{Z}_2$ symmetry is broken down to average. This statement is synonymous with the statement that, with only average $\mathbb{Z}_2$ symmetry, there is no nontrivial bosonic ASPT in $(2+1)d$. In particular, we discuss more details in Sec \ref{Sec: dynamics} that both decoherence and dissipation can drive the Luttinger liquid in each plaquette toward a mixed state with short-range entanglement. This leads to the formation of an SRE bulk mixed state. 

Subsequently, we turn our attention to the hinge mode. As we mentioned above, the hinge mode, which is also the building block of the wire construction, corresponds to the edge state of a $(2+1)d$ fermionic Levin-Gu state with a topological index $\nu=2$ with symmetry actions specified in Eq. \eqref{block symmetry}. We can introduce decoherence processes that break down the $\mathbb{Z}_2$ symmetry to an average $\mathbb{Z}_2$ symmetry. However, according to the general classification paradigm in Ref. \cite{average_2023}, $(2+1)d$ fermionic Levin-Gu state with topological index $\nu=2$ is still a nontrivial ASPT if we break $\Z_2$ symmetry to an average symmetry by some decoherence channel and keep the fermion parity $\Z_2^f$ exact. Therefore, the hinge mode, which is an edge of the ASPT mentioned above, still carries an average anomaly and cannot be turned into an SRE mixed state. The stability of the hinge modes shows that the system is a nontrivial type-\1 intrinsic SASPT.

\subsubsection{Type-\2 intrinsic SASPT from average anomalous building blocks}
\label{Sec: type-2 intrinsic}
In this subsection, our focus is on constructing intrinsic SASPT states whose building blocks do not have a clean limit. In other words, each building block is the boundary of a certain intrinsic ASPT state. 

As an example, let us consider the coupled-wire model with two-foliated subsystem fermion parity symmetry $G_s = \mathbb{Z}_2^f$ and global time-reversal symmetry $G_g = \mathbb{Z}_2^T$ with $(\mathbb{Z}_2^T
)^2=1$. In our previous discussion of closed systems in Sec. \ref{Sec: closed coupled-wire}, each building block in Fig. \ref{fig1} is expected to be an edge mode of a $(2+1)d$ $\mathbb{Z}_2^f \times \mathbb{Z}_2^T$ SPT state. However, the classification of $(2+1)d$ $\mathbb{Z}_2^f \times \mathbb{Z}_2^T$ SPT in pure states is trivial \cite{Wang_2020}, which means there are no nontrivial clean SSPT states in this symmetry class.

Now, let us consider open systems where decoherence breaks the time-reversal symmetry, transforming it into an average symmetry. In such settings, the classification of $(2+1)d$ ASPT with exact $\mathbb{Z}_2^f$ and average $\mathbb{Z}_2^T$ is actually nontrivial and all the nontrivial cases are intrinsic ASPTs (see Appendix \ref{App: intrinsic}). In this case, the coupled-wire model can be composed of the building block of $(1+1)d$ mixed states with exact fermion parity $\mathbb{Z}_2^f$ and average time-reversal $\mathbb{Z}_2^T$ symmetries which are the edges of the $(2+1)d$ intrinsic ASPTs. These building blocks cannot be SRE mixed states, indicating the presence of average anomalies. 

Here we provide more details on the classification data for the intrinsic ASPTs with exact $\mathbb{Z}_2^f$ and average $\mathbb{Z}_2^T$ symmetry. There are three layers of the classification data:
\begin{align}
\begin{gathered}
n_1\in\mathcal{H}^1[\Z_2^T,h^2(\Z_2^f)]=\Z_2\\
n_2\in\mathcal{H}^2[\Z_2^T,h^1(\Z_2^f)]=\Z_2\\
\nu_3\in\mathcal{H}^3[\Z_2^T,U(1)]=\Z_1
\end{gathered},
\end{align}
where $n_1$ labels the Majorana chain decoration on the time-reversal domain walls, $n_2$ labels the complex fermion decoration on the junctions of time-reversal domain walls, and $\nu_3$ represents the bosonic SPT state protected by $\mathbb{Z}_2^T$. In order to get a consistent clean SPT state, these data must satisfy the following consistency conditions (see Appendix \ref{APP: SS} for more details):
\begin{align}
\begin{aligned}
&\mathrm{d}_2n_1=s_1\cup n_1\cup n_1\\
&\mathrm{d}_2n_2=\mathcal{O}_4[n_2]
\end{aligned},
\end{align}
In the clean case, both $n_1$ and $n_2$ encounter obstructions, which is the reason for the absence of clean SPT states. However, in the context of open quantum systems, the $\mathcal{O}_4[n_2]$ obstruction is trivial due to the phase decoherence of open systems. Consequently, a nontrivial $n_2$ characterizes a $(2+1)d$ intrinsic fermionic ASPT phase. And this intrinsic ASPT has $\mathbb{Z}_2$ classification because nontrivial $n_2$ is actually the only intrinsic ASPT state for this symmetry class. The reason is nontrivial $n_1$ will lead to an obstruction of fermion parity violation, which is not allowed as long as $\mathbb{Z}_2^f$ is still an exact symmetry in the decohered systems \cite{average_2023}. As a result, we can set each building block depicted in Fig. \ref{fig1} to be the edge of the $(2+1)d$ intrinsic ASPT state with exact $\mathbb{Z}_2^f$ and average $\mathbb{Z}_2^T$ symmetry. 

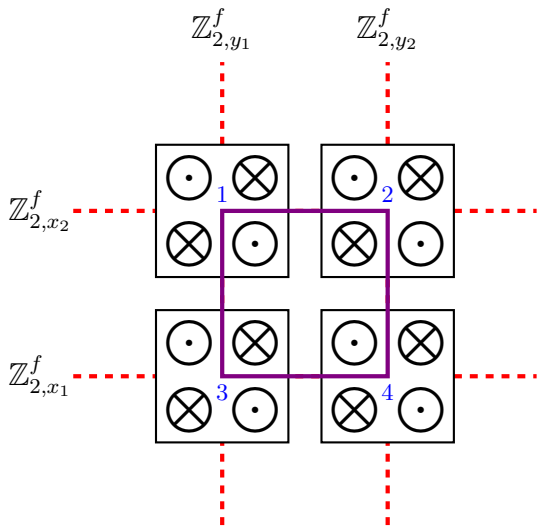
\begin{figure}
\centering
\begin{tikzpicture}[scale=0.88]
\tikzstyle{sergio}=[rectangle,draw=none]
\draw[dashed, ultra thick, color=red] (-2,0.25) -- (5,0.25);
\draw[dashed, ultra thick, color=red] (-2,2.75) -- (5,2.75);
\draw[dashed, ultra thick, color=red] (0.25,-2) -- (0.25,5);
\draw[dashed, ultra thick, color=red] (2.75,-2) -- (2.75,5);
\filldraw[thick, fill=white] (1.75,1.75) -- (1.75,3.75) -- (3.75,3.75) -- (3.75,1.75) -- cycle;
\filldraw[thick, fill=white] (1.25,1.75) -- (1.25,3.75) -- (-0.75,3.75) -- (-0.75,1.75) -- cycle;
\filldraw[thick, fill=white] (1.25,1.25) -- (1.25,-0.75) -- (-0.75,-0.75) -- (-0.75,1.25) -- cycle;
\filldraw[thick, fill=white] (1.75,1.25) -- (1.75,-0.75) -- (3.75,-0.75) -- (3.75,1.25) -- cycle;
\path (2.25,2.25) node [style=sergio] {\LARGE $\bigotimes$};
\path (2.25,3.25) node [style=sergio] {\LARGE $\bigodot$};
\path (3.25,3.25) node [style=sergio] {\LARGE $\bigotimes$};
\path (3.25,2.25) node [style=sergio] {\LARGE $\bigodot$};
\path (0.75,2.25) node [style=sergio] {\LARGE $\bigodot$};
\path (-0.25,2.25) node [style=sergio] {\LARGE $\bigotimes$};
\path (0.75,3.25) node [style=sergio] {\LARGE $\bigotimes$};
\path (-0.25,3.25) node [style=sergio] {\LARGE $\bigodot$};
\path (2.25,0.75) node [style=sergio] {\LARGE $\bigodot$};
\path (2.25,-0.25) node [style=sergio] {\LARGE $\bigotimes$};\path (3.25,0.75) node [style=sergio] {\LARGE $\bigotimes$};
\path (3.25,-0.25) node [style=sergio] {\LARGE $\bigodot$};
\path (0.75,0.75) node [style=sergio] {\LARGE $\bigotimes$};
\path (-0.25,0.75) node [style=sergio] {\LARGE $\bigodot$};
\path (0.75,-0.25) node [style=sergio] {\LARGE $\bigodot$};
\path (-0.25,-0.25) node [style=sergio] {\LARGE $\bigotimes$};
\path (-2.5,0.25) node [style=sergio] {\large $\Z_{2,x_1}^f$};
\path (-2.5,2.75) node [style=sergio] {\large $\Z_{2,x_2}^f$};
\path (0.25,5.5) node [style=sergio] {\large $\Z_{2,y_1}^f$};
\path (2.75,5.5) node [style=sergio] {\large $\Z_{2,y_2}^f$};
\path (2.75,3) node [style=sergio] {\small {\color{blue}2}};
\path (0.25,3) node [style=sergio] {\small {\color{blue}1}};
\path (2.75,0) node [style=sergio] {\small {\color{blue}4}};
\path (0.25,0) node [style=sergio] {\small {\color{blue}3}};
\draw[ultra thick, color=violet] (2.75,2.75) -- (0.25,2.75) -- (0.25,0.25) -- (2.75,0.25) -- cycle;
\end{tikzpicture}
\caption{Coupled-wire model of $(2+1)d$ type-\1 intrinsic SASPT state with exact subsystem fermion parity $\Z_{2,n}^f$ ($n=1,2,3,4$) and average global time-reversal $\Z_2^T$ symmetries. Four relevant conserved subsystem fermion parity $\Z_{2,n}^f$ ($n=1,2,3,4$) are remarked by red dashed lines. Each black solid block is a unit cell, and the purple dashed line defines a plaquette.}
\label{Z2f subsystem}
\end{figure}

In order to achieve an SRE bulk state, we need to argue that four building blocks from the four neighboring unit cells can be turned into an SRE ensemble using quantum decoherence (as shown in Fig. \ref{Z2f subsystem}). This is equivalent to saying that the total average anomaly vanishes for these modes. Examining the configuration, we observe the following charge assignments for the subsystem symmetries:
\begin{enumerate}[1.]
\item Wire-1 carries nontrivial charges associated with subsystem fermion parities $\mathbb{Z}_{2,x_2}^f$ and $\mathbb{Z}_{2,y_1}^f$.
\item Wire-2 carries nontrivial charges associated with subsystem fermion parity $\mathbb{Z}_{2,x_2}^f$ and $\mathbb{Z}_{2,y_2}^f$.
\item Wire-3 carries nontrivial charges associated with the subsystem fermion parity $\mathbb{Z}_{2,y_1}^f$ and $\mathbb{Z}_{2,x_1}^f$.
\item Wire-4 does not carry any charges associated with the subsystem fermion parities $\mathbb{Z}_{2,x_2}^f$ and $\mathbb{Z}_{2,y_1}^f$.
\end{enumerate}

To obtain an SRE ensemble from these four building blocks is to show that the average anomaly associated with all the symmetries is trivial. Let us consider the two subsystem fermion parities $\mathbb{Z}_{2,x_2}^f$ and $\mathbb{Z}_{2,y_1}^f$, we can summarize the average anomalies of different wires as follows:
\begin{align}
\begin{aligned}
&\text{wire-1: }n_2^{x_2}(g_1,g_2)+n_2^{y_1}(g_1,g_2)\\
&\text{wire-2: }n_2^{x_2}(g_1,g_2)\\
&\text{wire-3: }n_2^{y_1}(g_1,g_2)\\
&\text{wire-4: }0
\end{aligned},
\end{align}
where $n_2^{x_2}(g_1,g_2)\in\mathcal{H}^2[\Z_2^T,h^1(\Z_{2,x_2}^f)]=\Z_2$ and $n_2^{y_1}(g_1,g_2)\in\mathcal{H}^2[\Z_2^T,h^1(\Z_{2,y_1}^f)]=\Z_2$. Due to the $\mathbb{Z}_2$ nature of these anomalies, it is easy to see that the anomaly associated with these two symmetries is vanishing when we take all four wires into consideration. Similarly, one can easily check the total average anomaly of the four building blocks is trivial, guaranteeing the existence of an SRE bulk ensemble. Given the SRE bulk, from a similar construction as before, one can easily see that the hinge is a nontrivial mixed state which carries precisely the average anomaly of the edge of the $(2+1)d$ intrinsic ASPT state with exact $\mathbb{Z}_2^f$ and average $\mathbb{Z}_2^T$ symmetry. Thus, this system is a nontrivial intrinsic SASPT. 

A more explicit coupled-wire model (with some assumptions on the time reversal action) of this type-\2 intrinsic SASPT state is given in Appendix \ref{App: type-2} where we have used the doubled Hilbert space language to demonstrate this nontrivial state. In particular, the explicit decoherence channels needed to construct this state are provided in Appendix \ref{App: type-2}. 

\section{Classification scheme for the SASPT phases}
\label{Sec: classification}
In the previous section, we have demonstrated a few possibilities of SASPT phases in $(3+1)d$ open quantum systems with two-foliated subsystem symmetries. In particular, we discussed two distinct types of intrinsic SASPT states that cannot be realized as SSPT states in clean or closed quantum systems. In this section, we focus on the general classification scheme for SASPT phases. This classification scheme relies on the cancellation condition of average anomaly, similar to Eq. (\ref{two-foliated anomaly free}), within the framework of coupled-wire constructions.

\subsection{Classification by anomaly cancellation}
\label{Sec: average anomaly cancellation}
We discuss the general classification paradigm of SASPT phases based on the cancellation condition of an average anomaly in the coupled-wire models. To discuss the average anomaly, we recall the classification of ASPT with onsite symmetry. According to Ref. \cite{average_2023}, we know that, for onsite symmetry, the $\tilde{G}$-symmetric decohered ASPT density matrices $\rho$ are classified by the AH spectral sequence with modified data and differentials. The topological invariant of $\rho$ in $(d+1)d$ is a $(p+q)$-cocycle as an element of the $E_2$ page of the AH spectral sequence, as
\begin{align}
\nu_{p+q}(\{g\};\{k\})\in E_2^{p,q}=\H^p\left[G,h^q(K)\right],
\label{E2 page}
\end{align}
where $p+q=d+1$, $q>0$, and $(g_i, k_j)\in(G, K)$ are group elements of average and exact symmetries, respectively. Each element in Eq. (\ref{E2 page}) labels the average anomaly of a $d$-dimensional mixed state which is the boundary of the $(d+1)$-dimensional ASPT state. These average anomalous boundary systems will be our building blocks for the SASPT states. We note that our scheme is also applicable to fermionic systems, of which the fermion parity $\Z_2^f$ should be included as a subgroup of $K$.

Towards an SRE bulk state, the total average anomaly per plaquette (see Figs. \ref{fig1} and \ref{Z2f subsystem}) should be canceled. Therefore, similar to Eq. (\ref{two-foliated anomaly free}), for a two-foliated subsystem average symmetry $\tilde{G}_s$ that is extended from an exact subsystem symmetry $K_s$ and an average subsystem symmetry $G_s$, the anomaly cancellation map $\tilde{f}_2$ is modified to be
\begin{align}
\tilde{f}_2(\nu)(\{g,g'\};\{k,k'\})=\frac{\nu(\{g\};\{k\})\nu(\{g'\};\{k'\})}{\nu(\{gg'\};\{kk'\})}.
\label{two-foliated average anomaly cancellation}
\end{align}
To have average anomaly cancellation, we need the image of this map to be trivial. We note that there is an important difference between the average anomaly cancellation condition through $\tilde{f}_2$ [cf. Eq. (\ref{two-foliated average anomaly cancellation})] and that of the $f_2$ map [cf Eq. (\ref{two-foliated anomaly free})]. $\nu$ is the collection of all average obstruction-free elements in $E_2^{p,q}=\H^p[G_s, h^q(K_s)]$ with $q>0$, as the topological invariant of $(2+1)d$ ASPT phases. The total average anomaly cancellation condition is that image of the $\tilde{f}_2$ map falls into the trivial elements in $E_2^{p,q}=\H^p[G_s^2, h^q(K_s^2)]$ with $q>0$ or any element in $\H^{p+q}[G_s^2, U(1)]$. This is because there are no nontrivial ASPT states if there is no exact symmetry. 

\subsection{Case for intrinsic SASPT phases}
In Sec. \ref{Sec: average anomaly cancellation}, we established the cancellation of the average anomaly as the classification principle for SASPT phases, as expressed by Eq. (\ref{two-foliated average anomaly cancellation}). This cancellation bears a resemblance to the anomaly cancellation observed in closed systems. However, there is a key distinction between the $f_2$ map (\ref{two-foliated anomaly free}) and the $\tilde{f}_2$ map (\ref{two-foliated average anomaly cancellation}) in terms of their preimage and image groups, which leads to different classifications and intrinsic SASPTs. In Secs. \ref{sec: type-1 intrinsic} and \ref{Sec: type-2 intrinsic}, we presented two examples of intrinsic SASPT phases: the type-\1 and type-\2 SASPT phases. In this section, our objective is to see how they fit into the general classification.

Type-\1 intrinsic SASPT phases are defined by building blocks that correspond to the $d$-dimensional edge state of a $(d+1)$-dimensional clean SPT state. However, the ``gapping" problem of the bulk needs the assistance of decoherence. In this case, although it is not possible to find a symmetric interaction that fully gaps out each plaquette in the clean systems, by introducing all possible symmetric interactions, the modes in each plaquette can be deformed to the edge state of a $(d+1)$-dimensional SPT state labeled by an element of $\mathcal{H}^3[G_s^2, U(1)]$. Under decoherence of the $G_s$ degrees of freedom, the modes in each plaquette can be decohered to an SRE $(1+1)d$ mixed state. Therefore, the classification of type-\1 intrinsic SASPT phases includes the elements in Eq. (\ref{E2 page}) that label the ASPT phases in $(2+1)d$ with clean limits, and under the $\tilde{f}_2$ map (\ref{two-foliated average anomaly cancellation}) these elements are mapped to elements in $\H^{p+q}[G_s^2, U(1)]$.

In the type-\2 intrinsic SASPT phases, each building block corresponds to the edge state of a higher dimensional intrinsic ASPT state \cite{average_2023}. Under the $\tilde{f}_2$ map (\ref{two-foliated average anomaly cancellation}), if building blocks are mapped to trivial elements in $E_2^{p,q}=\H^p[G_s^2, h^q(K_s^2)]$ ($q>0$) or any element in $\H^{p+q}[G_s^2, U(1)]$, one can find a consistent type-II intrinsic SASPT.

\section{Effects of decoherence or dissipation on the hinge modes}
\label{Sec: dynamics}
In the previous section, we have discussed the classification of nontrivial SASPTs. We noted the importance of exact symmetries without which there are no robust SASPT phases. This means that, when all the symmetry of the system is broken down to average by decoherence, the hinge modes in principle can become SRE mixed state. However, in some cases, it is possible that the hinge mode is stable in some weak decoherence regime and eventually becomes SRE after some critical decoherence strength. In other words, the hinge mode could go through a dynamical transition driven by decoherence or dissipation.

We will now focus on a specific example in $(3+1)d$ with a two-foliated $\Z_2$ subsystem symmetry. According to a study by Ref. \cite{2022arXiv221015596Z}, this system can host a clean SSPT state whose hinge mode is expected to exhibit characteristics of the edge state of the $(2+1)d$ Levin-Gu model \cite{Levin-Gu}. The hinge mode can be described by 
\begin{align}
\mathcal{S}_{\rm LG}[\theta,\phi]=&\int\mathrm{d}x\mathrm{d}\tau\frac{1}{4\pi}\left(\partial_x\theta\partial_\tau\phi+\partial_x\phi\partial_\tau\theta\right)\nonumber\\
&-\frac{1}{8\pi}\left[K_0(\partial_x\theta)^2+\frac{4}{K_0}(\partial_x\phi)^2\right],
\label{Levin-Gu Lagrangian}
\end{align}
where $K_0$ is the Luttinger parameter and $\Z_2$ symmetry acts as $(\theta,\phi)\mapsto(\theta+\pi,\phi+\pi)$. If we consider decoherence or dissipation that breaks the $\mathbb{Z}_2$ subsystem symmetry down to average, there is no nontrivial SASPT in this case, which implies that the hinge mode can become SRE mixed state under decoherence or dissipation. However, small decoherence can be relevant or irrelevant depending on the value of the Luttinger parameter, and there could be an interesting transition on the hinge. 

A good way to handle the dynamical phase transition is to use the Choi-Jamiołkowski isomorphism, which maps the density matrix of the $(1+1)d$ wire to a pure state in the doubled Hilbert space, denoted by $\mathcal{H}_d$. Within the doubled space, we can transform the dynamical phase transition of the density matrix to the quantum phase transition of clean systems with some additional care.

\subsection{Effects of measurement/decoherence}
\label{Sec: measurement}

We use the Choi–Jamiołkowski isomorphism to investigate the influence of measurement/decoherence on the hinge state. We consider the quantum channels which are (weak) measurement/decoherence that breaks the $\mathbb{Z}_2$ exact symmetry down to average. We use renormalization group (RG) analysis in the doubled Hilbert space to study the effect of such measurement/decoherence. By Choi–Jamiołkowski isomorphism, the free part of the hinge theory in doubled Hilbert space is composed of two copies of Levin-Gu edge theory \eqref{Levin-Gu Lagrangian} which reads
\begin{align}
\mathcal{S}_{\rm LG}^l[\theta_l,\phi_l]-\mathcal{S}_{\rm LG}^r[\theta_r,\phi_r].
\label{double Levin-Gu edge}
\end{align}
We can view this theory as generated by a doubled space path integral in imaginary time. 

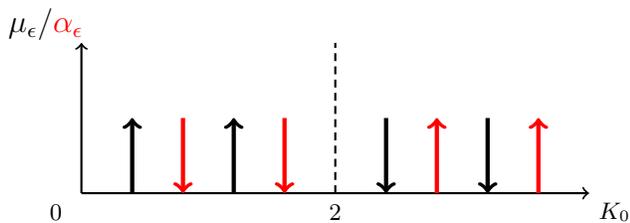
\begin{figure}
\begin{tikzpicture}[xscale=1.35, yscale=1]
\tikzstyle{sergio}=[rectangle,draw=none]
\draw[thick,->] (0,0)--(0,2);
\draw[thick, densely dashed] (2.5,0)--(2.5,2);
\draw[thick,->] (0,0)--(5,0);
\path (-0.25,-0.25) node [style=sergio]{$0$};
\path (5.25,-0.25) node [style=sergio]{$K_0$};
\path (2.5,-0.25) node [style=sergio]{$2$};
\path (-0.35,2.25) node [style=sergio]{\large $\mu_\epsilon$/{\color{red}$\alpha_\epsilon$}};
\draw[ultra thick, ->] (0.5,0)--(0.5,1);
\draw[ultra thick, ->] (1.5,0)--(1.5,1);
\draw[ultra thick, <-, red] (1,0)--(1,1);
\draw[ultra thick, <-, red] (2,0)--(2,1);
\draw[ultra thick, ->] (3,1)--(3,0);
\draw[ultra thick, ->] (4,1)--(4,0);
\draw[ultra thick, <-, red] (3.5,1)--(3.5,0);
\draw[ultra thick, <-, red] (4.5,1)--(4.5,0);
\end{tikzpicture}
\caption{RG flow phase diagram of the decoherence terms. The black and red arrows are RG flows of $\mu_\epsilon$ and $\alpha_\epsilon$ terms, respectively.}
\label{decoherence RG}
\end{figure}

At low energy, the weak measurement/decoherence which breaks the $\mathbb{Z}_2$ symmetry to average will be mapped to a coupling between $l$ and $r$ degrees of freedom at a given time \cite{Ehud_2022, JYLee_2022}. The simplest form of such coupling in low energy can be written as
\begin{align}
\begin{gathered}
S_{\phi}[\varphi_l,\varphi_r]=-\int\mathrm{d}x\sum\limits_{\epsilon=\pm}\mu_\epsilon\cos(\varphi_l+\epsilon\varphi_r)\\
S_{\theta}[\vartheta_l,\vartheta_r]=-\int\mathrm{d}x\sum\limits_{\epsilon=\pm}\alpha_\epsilon\cos(\vartheta_l+\epsilon\vartheta_r)
\end{gathered},
\label{weak measurements}
\end{align}
where $\varphi_{l/r}(x)=\phi_{l/r}(x,\tau=0)$ and $\vartheta_{l/r}(x)=\theta_{l/r}(x,\tau=0)$. It is easy to see that $S_\phi$ and $S_\theta$ break the $\Z_{2,l}$ and $\Z_{2,r}$ symmetries to the diagonal $\Z_{2}$ symmetry acting on $\H_l$ and $\H_r$ identically. In practice, there will be other kinds of perturbation that break the exact symmetry down to average. The above terms are conceivably the most relevant terms that can be generated by Kraus operators. 

Similar to the analysis in Ref. \cite{Ehud_2022, JYLee_2022, average_2023}, we perform a Wick rotation that exchanges the spatial coordinate $x$ with the imaginary time coordinate $\tau$. In the new coordinates, the weak measurements on \eqref{weak measurements} can be interpreted as a local coupling at $x=0$ that is constant along the imaginary time direction. The renormalization group (RG) equations for $S_{\phi}$ and $S_{\theta}$ take the form of RG equations for $(0+1)d$ static impurities in Luttinger liquids, which can be described as follows ($\epsilon=\pm$):
\begin{align}
\begin{gathered}
\frac{\mathrm{d}\mu_\epsilon}{\mathrm{d}l}=\left(1-\frac{K_0}{2}\right)\mu_\epsilon\\
\frac{\mathrm{d}\alpha_\epsilon}{\mathrm{d}l}=\left(1-\frac{2}{K_0}\right)\alpha_\epsilon\\
\frac{\mathrm{d}K_0}{\mathrm{d}l}=0
\end{gathered}.
\label{measurements}
\end{align}

As illustrated in Fig. \ref{decoherence RG}, for $K_0<2$, $S_{\phi}$ is relevant, otherwise $S_{\theta}$ is relevant in the infrared (IR) limit. When $S_{\phi}$ is relevant (namely $K_0<2$), one can show that the power-law correlation of $\langle e^{i\theta(r_1)}e^{-i\theta(r_2)}\rangle$ is spoiled. However, since the $\phi$ fields are commuting with the perturbation $S_\phi$, the correlation function $\langle e^{i\phi(r_1)}e^{-i\phi(r_2)}\rangle$ remains the same as before measurement/decoherence, namely power-law decay. As for $K_0>2$, the situation is opposite, namely $e^{i\theta}$ operators show nontrivial correlations. Therefore, in the weak measurement/decoherence regime, the nontrivial correlation on the hinge persists for any value of the Luttinger parameter $K_0$. However, we emphasize that the above are only analyses in the weak measurement/decoherence limit. If both terms in Eq. \eqref{measurements} are strong (say we only take the $\epsilon=+1$ terms), then they can pin the $\phi$ and $\theta$ fields and their connected correlations become trivial, which is consistent with the statement of a trivial bulk.

\subsection{Effects of dissipation}
\label{Sec: dissipation}
To study the effect of dissipation, we employ the Keldysh path integral formalism. In Keldysh, the Lindbladian $\L$ is expressed as
\begin{align}
&\L=-i(H_l-H_r)\\
+&\sum\limits_\alpha\gamma_\alpha\left[L_{\alpha,l}L_{\alpha,r}^*-\frac{1}{2}(L_{\alpha,l}^*L_{\alpha,l}+L_{\alpha,r}^*L_{\alpha,r})\right]\nonumber,
\label{Lindbladian}
\end{align}
where $H_{l/r}$ describe the unitary dynamics and $L_{l,r}$ contains the effects of dissipation. 

In our case, the unitary dynamics for the hinge mode is given by Eq. \eqref{Levin-Gu Lagrangian} (with imaginary time switched back to real time). The dissipation process in the low energy effective theory is expressed as couplings between the quantum fields on the left and right branches, which break the two individual $\mathbb{Z}_2$ symmetry on the left and right branches to a diagonal $\mathbb{Z}_2$ symmetry. We again consider the simplest possibility as the following
\begin{align}
\begin{gathered}
\L_\phi=-\sum\limits_{\epsilon=\pm}\mu_\epsilon\cos(\phi_l+\epsilon\phi_r)\\
\L_\theta=-\sum\limits_{\epsilon=\pm}\alpha_\epsilon\cos(\theta_l+\epsilon\theta_r)
\end{gathered}.
\end{align}
Note that these terms do not come with a $\delta$ function in time, i.e., the dissipation process happens continuously in time. We can regard these terms as coming from continuous weak measurements done by the environment on the system. 
\begin{figure}
\begin{tikzpicture}[xscale=1.35, yscale=1]
\tikzstyle{sergio}=[rectangle,draw=none]
\draw[thick,->] (0,0)--(0,2);
\draw[thick, densely dashed] (1,0)--(1,2);
\draw[thick, densely dashed] (4,0)--(4,2);
\draw[thick,->] (0,0)--(5,0);
\path (-0.25,-0.25) node [style=sergio]{$0$};
\path (5.25,-0.25) node [style=sergio]{$K_0$};
\path (1,-0.25) node [style=sergio]{$1$};
\path (4,-0.25) node [style=sergio]{$4$};
\path (-0.35,2.25) node [style=sergio]{\large $\mu_\epsilon$/{\color{red}$\alpha_\epsilon$}};
\draw[ultra thick, ->] (0.33,0)--(0.33,1);
\draw[ultra thick, ->, color=red] (0.66,1)--(0.66,0);
\draw[ultra thick, ->] (1.33,0)--(1.33,1);
\draw[ultra thick, ->, color=red] (1.66,0)--(1.66,1);
\draw[ultra thick, ->] (2,0)--(2,1);
\draw[ultra thick, ->, color=red] (2.33,0)--(2.33,1);
\draw[ultra thick, ->] (2.66,0)--(2.66,1);
\draw[ultra thick, ->, color=red] (3,0)--(3,1);
\draw[ultra thick, ->] (3.33,0)--(3.33,1);
\draw[ultra thick, ->, color=red] (3.66,0)--(3.66,1);
\draw[ultra thick, ->] (4.33,1)--(4.33,0);
\draw[ultra thick, ->, color=red] (4.66,0)--(4.66,1);
\end{tikzpicture}
\caption{RG flow phase diagram of the dissipation terms. There are three different regimes $K_0<1$, $1<K_0<4$, and $K_0>4$ where the relevance of the two terms at small coupling are demonstrated. The black and red arrows are RG flows of $\mu_\epsilon$ and $\alpha_\epsilon$, respectively.}
\label{dissipation RG}
\end{figure}
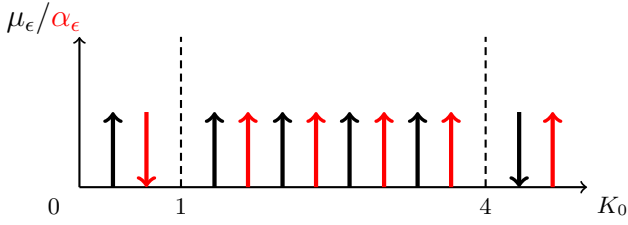

It is easy to verify that the scaling dimensions of the ratios $\mu_-/\mu_+$ and $\alpha_-/\alpha_+$ vanish, namely, $\mu_-/\mu_+$ and $\alpha_-/\alpha_+$ do not flow under the renormalization group. To simplify our analysis, we choose a submanifold in the parameter space with $\mu_-=\alpha_-=0$. It is easy to see that this submanifold is closed under RG flow. The RG equations for the coupling constants $\mu_+$, $\alpha_+$, and the Luttinger parameter $K_0$ are given by \cite{Giamarchi_2003}
\begin{align}
\begin{gathered}
\frac{\mathrm{d}\mu_+}{\mathrm{d}l}=\left(2-\frac{K_0}{2}\right)\mu_+\\
\frac{\mathrm{d}\alpha_+}{\mathrm{d}l}=\left(2-\frac{2}{K_0}\right)\alpha_+\\
\frac{\mathrm{d}K_0}{\mathrm{d}l}=-\frac{1}{4}\alpha_+^2+\frac{1}{16}K_0^2\mu_+^2
\end{gathered}
~.
\end{align}

Notice the RG flow for the Luttinger parameter differs by a sign from the usual RG equation in the clean systems. This is due to the non-Hermitian nature of the dissipation terms. It is hard to determine the full flow of the RG equation. Due to the perturbative nature of the RG equations, we will treat $K$ as a free parameter and only look at the relevance of the $\mu_+$ and $\alpha_+$ terms. There are three different regimes as shown in Fig. \ref{dissipation RG},
\begin{enumerate}[1.]
\item $K_0<1$: $\L_\phi$ is relevant while $\L_\theta$ is irrelevant. This region corresponds to a mixed hinge state with the power-law correlation of $\langle e^{i\phi_{l/r}(r_1)}e^{-i\phi_{l/r}(r_2)}\rangle$ (because the dissipation commutes with the operator $e^{-i\phi_{l/r}(r)}$) and short-range correlation of $\langle e^{i\theta_{l/r}(r_1)}e^{-i\theta_{l/m}(r_2)}\rangle$.
\item $1<K_0<4$: both $\L_\phi$ and $\L_\theta$ are relevant. This region corresponds to a mixed state with short-range correlation for all operators, i.e. a trivial state.
\item $K_0>4$: $\L_\theta$ is relevant while $\L_\phi$ is irrelevant. This region corresponds to a mixed hinge state with the power-law correlation of $\langle e^{i\theta_{l/r}(r_1)}e^{-i\theta_{l/r}(r_2)}\rangle$ and short-range correlation of $\langle e^{i\phi_{l/r}(r_1)}e^{-i\phi_{l/r}(r_2)}\rangle$.
\end{enumerate}

Therefore, in the intermediate regime of the phase diagram, the power-law correlations of the hinge mode are destroyed by the dissipation.

\section{Conclusion and outlook}
\label{Sec: Summary}
In this work, we investigate the construction and classification of fractonic higher-order topological phases in open quantum systems with exact and average symmetries. Our analysis focuses on scenarios of $(3+1)d$ systems with a two-foliated subsystem symmetry. In particular, we demonstrate two types of intrinsic SASPT phases that cannot exist in the pure state. For type-\1 intrinsic SASPT phases, a trivial atomic insulator exists in the pure state, however, the existence of the nontrivial SPT state requires the help of quantum decoherence. For type-\2 intrinsic SASPT, in some sense, even the existence of the trivial atomic insulator requires decoherence. Throughout the discussion, the notion of average anomaly (which describes the boundary of an average SPT state) plays an important role and we use this idea to provide a general classification of these SASPT states. It is worth noting that the average anomaly-free condition for SASPT phases deviates slightly from that of SSPT phases in closed (clean) systems. Specifically, we consider all elements in $H^{d+1}[G_s^2, U(1)]$ to be trivial, since average anomalies do not arise in systems with average symmetry only \cite{MaWangASPT, average_2023}. We also study the effects of measurement or dissipation on the hinge modes of fractonic higher-order topological phases. Examples of interesting dynamical phase transitions are discussed. 

The investigation of symmetry and topology in the presence of decoherence and dissipation holds significant importance and interest in the interdisciplinary realm of quantum many-body physics and quantum information. The insights gained from studying the modification of consistency conditions for average symmetry in dissipative foliated systems can be extended to systems with higher-form symmetries and intrinsic topological orders. Furthermore, in addition to decoherence and dissipation, the notion of average symmetry is also very relevant to disordered systems. However, fractonic phases with disorder could be much richer in terms of the dynamical properties of their excitations. On one hand, disorders can have the effect of generating mobility for the fractonic excitations since the disorder breaks subsystem symmetries. On the other hand, with strong disorder, systems are also expected to be localized near the ground state. These competing effects indicate that phase diagrams of fractonic phases as a function of symmetry-breaking disorder strength can be very complex and worth a careful study in the future. 

\begin{acknowledgements}
We thank Ruochen Ma, Yichen Xu, and Chong Wang for stimulating discussions and Meng Cheng for a previous collaboration. J.H.Z. and Z.B. are supported by a startup fund from the Pennsylvania State University and thank the hospitality of the Kavli Institute for Theoretical Physics, which is partially supported by the National Science Foundation under Grant No. NSF PHY-1748958. J.H.Z. appreciates the hospitality of Zheng-Cheng Gu at the Chinese University of Hong Kong, Z.B. is also grateful to the hospitality of the Institute for Advanced Study at Tsinghua University. K.D. and S.Y. are supported by the National Natural Science Foundation of China (NSFC) (Grant No. 12174214 and No. 92065205), the National Key R\&D Program of China (Grant No. 2018YFA0306504), the Innovation Program for Quantum Science and Technology (Grant No. 2021ZD0302100), and the Tsinghua University Initiative Scientific Research Program.
\end{acknowledgements}

\appendix
\setcounter{equation}{0}
\renewcommand{\theequation}{S%
\arabic{equation}} \setcounter{equation}{0} \renewcommand{\thefigure}{S%
\arabic{figure}} \setcounter{figure}{0}

\section{A brief review of ASPT classification}
\label{APP: SS}
The formal classification of ASPT phases with global symmetry can be done through the generalized spectral sequence method. Consider the general group extension,
\begin{equation}
    1\rightarrow K\rightarrow\tilde{G}\rightarrow G\rightarrow1,
\label{app extension}
\end{equation}
with $G$ being the average symmetry group and $K$ being the exact symmetry group (for fermionic systems, the fermion parity $\Z_2^f$ should be included as a subgroup of $K$). Mathematically, the consistency conditions for decorated domain walls with symmetry $\tilde{G}$ are consolidated into an Atiyah-Hirzebruch (AH) spectral sequence. All possible decorated domain wall patterns are summarized as the so-called $E_2$ page of this spectral sequence, and it is given by:
\begin{align}
\bigoplus\limits_{p+q=d+1}E_2^{p,q}=\bigoplus\limits_{p+q=d+1}H^p[G, h^q(K)].
\end{align}
In the above equation, $h^q(K)$ represents the classification of $K$-symmetric invertible topological phases in $q$ dimensions, which are decorated on the defects of $G$ symmetry with a codimension of $q$. It is important to note the following modifications which are distinct from the ordinary spectral sequence for classifying the SPT phases in the clean systems: 
\begin{enumerate}[1.]
\item $h^0(K)=0$ because there is no nontrivial ASPT state if there is no exact symmetry.
\item Bosonic invertible topological phases should be excluded from $h^q(K)$ (for example, $(2+1)d$ Kitaev's $E_8$ state is excluded) -- this is because such states can be prepared by a finite-depth quantum channel from a trivial product state \cite{MaWangASPT}. 
\end{enumerate}

As mentioned above, not all domain wall configurations can give rise to nontrivial SPT states, as certain consistency conditions need to be satisfied during the construction of an SPT wave function in clean systems. Within the framework of the AH spectral sequence, the consistency conditions are captured by the differentials denoted as $\mathrm{d}_r$, which map elements from $E_2^{p,q}$ to decorated domain wall configurations in $E_2^{p+r,q-r+1}$ in one higher dimension, namely
\begin{align}
\mathrm{d}_r:~E_{2}^{p,q}\rightarrow E_2^{p+r,q-r+1}.
\label{differential}
\end{align}
These consistency conditions ensure that the symmetry defect of $G$ symmetry can quantum fluctuate in a wavefunction while keeping SRE properties \cite{Wang_2021}. In particular, the final layer of the differential $\mathrm{d}_{q+1}$ ensures that no Berry phase is accumulated after a closed path of continuous domain wall deformation. For open quantum systems, the Berry phases of different decorated domain wall patterns are no longer well-defined, therefore, we abandon the Berry phase consistency condition for open quantum systems. Mathematically, we delete the last layer of obstruction $\mathrm{d}_{q+1}$ in Eq. \eqref{differential} when calculating the classification of ASPT phases in open quantum systems. 

For more details about the Atiyah-Hirzebruch (AH) and Lyndon-Hochschild-Serre (LHS) spectral sequences, see Refs. \cite{Wang_2020, Wang_2021, average_2023}. In the following, we present a simple example to sketch the spectral sequence of calculating the ASPT classification. 

\subsection{$\Z_2^T\times\Z_2^f$ fSPT phases in $(2+1)d$}
\label{App: intrinsic}
In this section, we give an example of classification of the $(2+1)d$ fSPT phases with $\Z_2^T\times\Z_2^f$ by AH spectral sequence \cite{Wang_2018, Wang_2020, Wang_2021}. We list all possible classification data as the elements of $E_2$-page of AH spectral sequence, as
\begin{enumerate}[1.]
\item $n_1\in E_2^{1,2}=H^1[\Z_2^T, h^2(\Z_2^f)]=\Z_2$: Kitaev's Majorana chain decoration on the codimension-1 $\Z_2^T$ domain wall, where $h^2(\Z_2^f)=\Z_2$ is the classification of $(1+1)d$ Kitaev's Majorana chain.
\item $n_2\in E_2^{2,1}=H^2[\Z_2^T, h^1(\Z_2^f)]=\Z_2$: Complex fermion decoration on the codimension-2 $\Z_2^T$ domain wall junction, where $h^1(\Z_2^f)=\Z_2$ is the classification of complex fermion parity.
\item $\nu_3\in E_2^{3,0}=H^3[\Z_2^T, U(1)]=\Z_1$.
\end{enumerate}
And the differentials are defined as
\begin{align}
\begin{gathered}
\mathrm{d}_2n_1=s_1\cup n_1\cup n_1\\
\mathrm{d}_3n_1=0\\
\mathrm{d}_2n_2=\mathcal{O}_4[n_2]
\end{gathered},
\end{align}
where $s_1$ characterizes the anti-unitary elements, such that 
\begin{align}
s_1(g)=\left\{
\begin{aligned}
&1,~~\text{$g$ is anti-unitary}\\
&0,~~\text{$g$ is unitary}
\end{aligned}
\right.,
\end{align}
and $\mathcal{O}_4[n_2]$ is the obstruction function of the Berry phase consistency, as a function of $n_2$ \cite{Wang_2020}. In clean systems, all possible decorated domain wall patterns are obstructed, towards a trivial classification. In open quantum systems with an average $\Z_2^T$ symmetry and exact $\Z_2^f$ symmetry, the Berry phase consistency would be lifted because of the decoherence, as a consequence, the nontrivial $n_2$ data corresponds to an intrinsic ASPT state. On the other hand, $n_1$ is still obstructed, therefore, the eventual classification for ASPT in this symmetry class is $\mathbb{Z}_2$.  

\section{Choi–Jamiołkowski isomorphism}
\label{channel-state duality}
We describe some more details on the Choi–Jamiołkowski isomorphism and the mapping between ASPT density matrices and SPT states in the doubled space. An ASPT density matrix, denoted by
\begin{equation}
    \rho=\sum_jp_j|\psi_j\rangle\langle\psi_j|,
    \label{density matrix}
\end{equation}
defined on the Hilbert space $\H$ will be mapped to the following Choi state in the doubled Hilbert space $\H_d=\H_l\otimes\H_r$ under the Choi–Jamiołkowski isomorphism,
\begin{align}
|\rho\rrangle=\frac{1}{\sqrt{\mathrm{dim}(\rho)}}\sum\limits_jp_j|\psi_j\rangle\otimes|\psi_j^*\rangle,
\label{Choi state}
\end{align}
where both left Hilbert space $\H_l$ and right Hilbert space $\H_r$ are identical with the physical Hilbert space $\H$. The doubled Hilbert space comes with a larger symmetry group.  

The exact symmetry $K$ is ``doubled" in the doubled space to $K_l\times K_r$ symmetry, namely 
\begin{align}
\begin{aligned}
&U_{k,l}|\rho\rrangle=\frac{1}{\sqrt{\mathrm{dim}(\rho)}}\sum\limits_jp_j\left(U_{k}|\psi_j\rangle\right)\otimes|\psi_j^*\rangle\\
&U_{k,r}|\rho\rrangle=\frac{1}{\sqrt{\mathrm{dim}(\rho)}}\sum\limits_jp_j|\psi_j\rangle\otimes\left(U_{k}^*|\psi_j^*\rangle\right)
\end{aligned}.
\end{align}
The average symmetry $G$ is mapped to $G_d$ symmetry in $\H_d$, such that
\begin{align}
U_{g,d}|\rho\rrangle=\frac{1}{\sqrt{\mathrm{dim}(\rho)}}\sum\limits_jp_j\left(U_g|\psi_j\rangle\right)\otimes\left(U_{g}^*|\psi_j^*\rangle\right).
\end{align}
There is also an anti-unitary ``SWAP$^*$'' symmetry as
\begin{align}
\mathrm{SWAP}^*\equiv\mathcal{C}\circ\mathrm{SWAP},
\end{align}
where $\mathcal{C}$ is the complex conjugation, and SWAP symmetry exchanges $\H_l$ and $\H_r$.

Therefore, a $\tilde{G}$-symmetric (\ref{app extension}) density matrix $\rho$ is mapped to a $\tilde{G}_d\rtimes\mathrm{SWAP}^*$-symmetric quantum state $|\rho\rrangle$, where $\tilde{G}_d$ is characterized by
\begin{align}
1\rightarrow K_l\times K_r\rightarrow\tilde{G}_d\rightarrow G_d\rightarrow1.
\label{double symmetry}
\end{align}

Measurements or decoherence can be in general described by some local quantum channels $\E$,
\begin{align}
\E[\rho]=\E_1\circ\E_2\circ\cdots\circ\E_{N}[ \rho ]=\sum\limits_{j=1,k}^NK_{j,k}\rho K_{j,k}^\dag,
\label{quantum channel}
\end{align}
where $K_{j,k}$'s are the local Kraus operators on site-$j$, satisfying the condition
\begin{align}
\sum\limits_kK_{j,k}^\dag K_{j,k}=1,~\forall j=1,\cdots,N.
\label{Kraus operators}
\end{align}
In the doubled Hilbert space $\mathcal{H}_d$, the quantum channel $\E$ (\ref{quantum channel}) in $\H$ is mapped to the following (non-unitary in general) operator in $\H_d$, 
\begin{align}
\hat{\E}_j=\sum\limits_kK_{k,l}\otimes K_{k,r}^*.
\end{align}

\begin{figure}
\centering
\begin{tikzpicture}[scale=0.88]
\tikzstyle{sergio}=[rectangle,draw=none]
\draw[dashed, ultra thick, color=DarkGreen] (-2.5,-0.625) -- (5.5,-0.625);
\draw[dashed, ultra thick, color=DarkGreen] (-2.5,2.625) -- (5.5,2.625);
\draw[dashed, ultra thick, color=DarkGreen] (-0.125,-3) -- (-0.125,5);
\draw[dashed, ultra thick, color=DarkGreen] (3.125,-3) -- (3.125,5);
\filldraw[thick, fill=white] (1.75,1.25) -- (1.75,4) -- (4.5,4) -- (4.5,1.25) -- cycle;
\filldraw[thick, fill=white] (1.25,1.25) -- (1.25,4) -- (-1.5,4) -- (-1.5,1.25) -- cycle;
\filldraw[thick, fill=white] (1.25,0.75) -- (1.25,-2) -- (-1.5,-2) -- (-1.5,0.75) -- cycle;
\filldraw[thick, fill=white] (1.75,0.75) -- (1.75,-2) -- (4.5,-2) -- (4.5,0.75) -- cycle;
\path (2.25,2.25) node [style=sergio] {\Large $\bigotimes$};
\path (2.75,1.75) node [style=sergio] {\Large $\bigodot$};
\path (2.25,3) node [style=sergio] {\Large $\bigotimes$};
\path (2.75,3.5) node [style=sergio] {\Large $\bigodot$};
\path (3.5,3.5) node [style=sergio] {\Large $\bigodot$};
\path (4,3) node [style=sergio] {\Large $\bigotimes$};
\path (3.5,1.75) node [style=sergio] {\Large $\bigodot$};
\path (4,2.25) node [style=sergio] {\Large $\bigotimes$};
\path (0.75,2.25) node [style=sergio] {\Large $\bigotimes$};
\path (0.25,1.75) node [style=sergio] {\Large $\bigodot$};
\path (-0.5,1.75) node [style=sergio] {\Large $\bigodot$};
\path (-1,2.25) node [style=sergio] {\Large $\bigotimes$};
\path (-1,3) node [style=sergio] {\Large $\bigotimes$};
\path (0.75,3) node [style=sergio] {\Large $\bigotimes$};
\path (-0.5,3.5) node [style=sergio] {\Large $\bigodot$};
\path (0.25,3.5) node [style=sergio] {\Large $\bigodot$};
\path (0.25,0.25) node [style=sergio] {\Large $\bigodot$};
\path (0.75,-0.25) node [style=sergio] {\Large $\bigotimes$};
\path (0.75,-1) node [style=sergio] {\Large $\bigotimes$};
\path (0.25,-1.5) node [style=sergio] {\Large $\bigodot$};
\path (-0.5,-1.5) node [style=sergio] {\Large $\bigodot$};
\path (-1,-1) node [style=sergio] {\Large $\bigotimes$};
\path (-1,-0.25) node [style=sergio] {\Large $\bigotimes$};
\path (-0.5,0.25) node [style=sergio] {\Large $\bigodot$};
\path (0.25+3.25,0.25) node [style=sergio] {\Large $\bigodot$};
\path (0.75+3.25,-0.25) node [style=sergio] {\Large $\bigotimes$};
\path (0.75+3.25,-1) node [style=sergio] {\Large $\bigotimes$};
\path (0.25+3.25,-1.5) node [style=sergio] {\Large $\bigodot$};
\path (-0.5+3.25,-1.5) node [style=sergio] {\Large $\bigodot$};
\path (-1+3.25,-1) node [style=sergio] {\Large $\bigotimes$};
\path (-1+3.25,-0.25) node [style=sergio] {\Large $\bigotimes$};
\path (-0.5+3.25,0.25) node [style=sergio] {\Large $\bigodot$};
\path (2,1.6) node [style=sergio] {\large{\color{blue}\2}};
\path (1,1.6) node [style=sergio] {\large{\color{blue}\1}};
\path (2.05,0.4) node [style=sergio] {\large{\color{blue}\4}};
\path (0.95,0.4) node [style=sergio] {\large{\color{blue}\3}};
\draw[ultra thick, color=violet] (3.125,2.625) -- (-0.125,2.625) -- (-0.125,-0.625) -- (3.125,-0.625) -- cycle;
\draw[draw=red, thick, rotate around={45:(2.5,0)}] (2.5,0) ellipse (20pt and 11pt);
\draw[draw=red, thick, rotate around={315:(0.5,0)}] (0.5,0) ellipse (20pt and 11pt);
\draw[draw=red, thick, rotate around={315:(2.5,2)}] (2.5,2) ellipse (20pt and 11pt);
\draw[draw=red, thick, rotate around={45:(0.5,2)}] (0.5,2) ellipse (20pt and 11pt);
\end{tikzpicture}
\caption{Coupled-wire construction of SASPT states in doubled Hilbert space $\H_d$. Each red ellipse includes one ``$\bigotimes$'' and one ``$\bigodot$'' which depict the edge theory of an ASPT state $|\rho\rrangle$ in doubled Hilbert space $\H_d$.}
\label{fig_ave}
\end{figure}
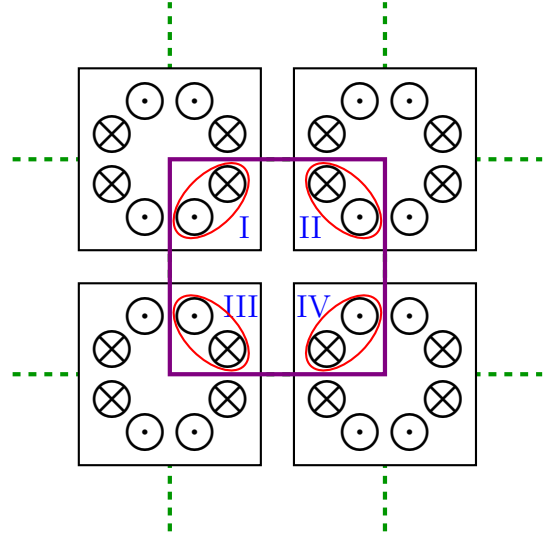

Physically, the quantum channels in the doubled space act as an interaction between the left and right space. Specifically, the interactions are symmetric under $K_l\times K_r$ but break $G_l\times G_r$ down to the diagonal $G_d$ symmetry. With the Choi–Jamiołkowski isomorphism, we are ready to formulate the description of average anomalies as the 't Hooft anomalies of enlarged symmetry in the doubled Hilbert space $\H_d$. Suppose the density matrix in Eq. \ref{density matrix} describes an ASPT state, then each component of the density matrix is described by a certain consistent domain wall decoration pattern which is labeled by $\nu_{p+q}(g_1,\cdots,g_p;k_{p+1},\cdots,k_{p+q})$. It can be shown that the Choi state $|\rho\rrangle$ of the density matrix $\rho$ is an SPT wavefunction in $\H_d$, with the following topological invariant
\begin{align}
\omega_{p+q}\coloneqq\frac{\nu_{p+q}(g_1,\cdots,g_p;k_{p+1,l},\cdots,k_{p+q,l})}
{\nu_{p+q}(g_1,\cdots,g_p;k_{p+1,r},\cdots,k_{p+q,r})}.
\label{double anomaly}
\end{align}
This topological invariant can be demonstrated to be a cocycle in $\H^p[G_d, h^q(K_l\times K_r)]$, where $k_{i,l}\in K_l$, $k_{i,r}\in K_r$ and $g_j\in G_d$. Hence, it represents an SPT wavefunction in the doubled space. 

In particular, this mapping also encompasses the intrinsic ASPTs -- we can demonstrate that the density matrix of an intrinsic ASPT state will also be mapped to an SPT state in the doubled Hilbert space. One can understand this by the following arguments. Consider the topological invariant of a $\tilde{G}$-symmetric intrinsic ASPT state $\nu_{p+q}\in E_2^{p,q}$, which under the differential $\mathrm{d}_{q+1}$ is mapped to a nontrivial element $\nu_{p+q+1}\in E_2^{p+q+1,0}$ in one higher dimension -- meaning this particular decorated domain wall pattern is obstructed in a clean system. This obstruction is described by a nontrivial cocycle $\nu_{p+q+1}\in H^{p+q+1}[G, U(1)]$ which indicates an inconsistent Berry phase accumulation along a closed path of deformation of $G$ domain walls. We know such a decoration pattern can be consistent in the mixed state and it correponds to an intrinsic ASPT state. On the other hand, for the corresponding Choi state, the total Berry phase would be $\nu_{p+q+1}\nu_{p+q+1}^*$ which can be shown to automatically fall into the trivial element in $H^{p+q+1}[G, U(1)]$. Hence the Choi state $|\rho\rrangle$ is obstruction-free in doubled Hilbert space. One can also show that the cocycle in Eq. \eqref{double anomaly} represents a nontrivial SPT in doubled space.

\section{Coupled-wire model of type-\2 intrinsic SASPT}
\label{App: type-2}
In this section, we present a conjecture of a coupled-wire model in doubled Hilbert space for the type-\2 intrinsic SASPT with exact subsystem $\mathbb{Z}_2^f$ symmetries and average time reversal symmetry $\mathbb{Z}_2^T$ given in Sec. \ref{Sec: coupled-wire}. 

In the wire construction, we use the building blocks which are the edge of an intrinsic ASPT with exact $\mathbb{Z}_2^f$ and average $\mathbb{Z}_2^T$ symmetry. Unfortunately, we currently do not have a first principle way to determine the boundary theory of an intrinsic ASPT. However, we know some requirements in the doubled space are needed for this construction. First of all, the exact fermion parity symmetry for the left and right space must factorize. Second, since the theory is supposed to be the boundary of intrinsic ASPT, therefore, one should not be able to factorize the time reversal action into actions in each individual subspace. Otherwise, the mixed state will have a clean limit and hence not be intrinsic. Of course, the average time reversal symmetry should commute with the swap symmetry. The final requirement is that the theory in doubled space is anomalous given these symmetry assignments. With these requirements, there might not be a unique choice of edge theory. Nonetheless, in the following, we give one example of construction that satisfies the above requirements. We conjecture that this theory can be an edge theory of the intrinsic ASPT with exact $\mathbb{Z}_2^f$ and average $\mathbb{Z}_2^T$ symmetry. 

One such theory in the doubled space is a Luttinger liquid with a four-component boson field, with the $K$-matrix $K=\sigma_l^z\oplus\sigma_r^z$, where the two blocks correspond to the left and right spaces as displayed in Fig. \ref{fig_ave}. The fermion parities in the Hilbert spaces $\H_l$ and $\H_r$ spaces are uniquely given by
\begin{align}
\begin{gathered}
W^{P_f^l}=\mathbbm{1}_{4\times4},~\delta\phi^{P_f^l}=\pi(1,1,0,0)^T\\
W^{P_f^r}=\mathbbm{1}_{4\times4},~\delta\phi^{P_f^r}=\pi(0,0,1,1)^T
\end{gathered},
\end{align}
and the SWAP$^*$ symmetry is uniquely defined as
\begin{align}
W^\mathrm{S}=\begin{pmatrix}
0 & \sigma^x\\
\sigma^x & 0
\end{pmatrix},~\delta\phi^{\mathrm{S}}=0.
\label{swap}
\end{align}
The time-reversal symmetry transformation is tricky. We need a transformation matrix that is not factorizable in the left and right spaces. The transformation matrix should be commuting with $W^S$. And finally, the time-reversal symmetry should have an anomaly that manifests the decorated domain wall picture (i.e., a time-reversal domain wall decorated by two complex fermions, one is from the left Hilbert space and the other is from the right Hilbert space). By brute-force search, we find such a time reversal action that satisfies all these requirements as the following, 
\begin{align}
W^{\T}=\begin{pmatrix}
0 & 1 & -1 & -1\\
1 & 0 & 1 & 1\\
1 & 1 & 0 & 1\\
-1 & -1 & 1 & 0
\end{pmatrix},~\delta\phi^{\T}=\left(\begin{array}{cccc}
0\\
\pi\\
\pi\\
0
\end{array}\right).
\label{average time-reversal}
\end{align}

At the outset, one should check the $W^\T$ commute with the two fermion parities and the swap symmetry, and $\T^2=1$, which is consistent with our symmetry action assignment. The tricky part is to show the mixed anomaly between the time-reversal symmetry and the two fermion parities. First, one can show that there is no gapping term one can turn on to get rid of these modes without breaking symmetry either explicitly or spontaneously. In particular, any term with the following form is not compatible with the time-reversal symmetry \eqref{average time-reversal}:
\begin{align}
\cos(a\phi_1+b\phi_2+c\phi_3+d\phi_4+\varphi),~a,b,c,d\in\Z,~\varphi\in [0,2\pi)
\end{align}
This indicates that indeed these symmetries are anomalous. But to more precisely demonstrate the anomaly, one way to do it is to explicitly break the time-reversal symmetry by some order parameter and show that there are nontrivial fermion zero modes (one from the left sector and one from the right sector) localized at the domain wall of this order parameter. To that end, we can consider the following time-reversal order parameters, 
\begin{align}
H_{\mathrm{TB}}=m(x)\cos(\phi_1+\phi_2)+m(x)\cos(\phi_3+\phi_4),
\label{Cooper}
\end{align}
It is easy to see that \eqref{Cooper} is compatible with the SWAP$^*$ symmetry \eqref{swap} and the two fermion parities, however, it explicitly breaks time-reversal symmetry, namely
\begin{align}
\mathcal{T}:~\left\{\begin{aligned}
&\cos(\phi_1+\phi_2)\\
&\cos(\phi_3+\phi_4)
\end{aligned}\right.\mapsto
\left\{\begin{aligned}
&-\cos(\phi_1+\phi_2)\\
&-\cos(\phi_3+\phi_4)
\end{aligned}\right.
.
\end{align}
We can see these backscattering terms are Cooper pair terms by re-fermionization. If we make a domain wall configuration of $m$, standard calculation can explicitly show that at the time-reversal symmetry domain wall, each sector has exactly one complex fermion zero mode. Therefore, the symmetry assignment indeed carries the anomaly we want to study. So far we have constructed a reasonable conjecture for the theory in each building block in the doubled space. We note that this construction might not be unique. However, from the $\mathbb{Z}_2$ classification of the ASPT state, one can infer that all possible constructions for the nontrivial state are equivalent in the sense of average anomaly. 

Now we can come to the question of constructing an SASPT state using these building blocks. To that end, as usual, we should introduce symmetric gapping terms in each plaquette to get a symmetric gapped bulk, the only difference is now the construction is in the doubled space. In each plaquette, there are 16 bosonic modes in the doubled Hilbert space whose $K$-matrix is given by $\widetilde{K}=K\oplus-K\oplus K\oplus-K$. We can find eight null vectors and keep both the average time-reversal and the double subsystems fermion parity $\mathbb{Z}_{2,n}^f$ as well as the swap symmetry. These symmetric Higgs terms read
\begin{align}
\mathcal{L}_{\mathrm{Higgs}}=&\cos\left(\phi_1^{\1}+\phi_4^{\1}+\phi_1^{\2}+\phi_4^{\2}\right)\nonumber\\
&+\cos\left(\phi_1^{\1}+\phi_4^{\1}+\phi_1^{\3}+\phi_4^{\3}\right)\nonumber\\
&+\cos\left(\phi_1^{\3}+\phi_4^{\3}+\phi_1^{\4}+\phi_4^{\4}\right)\nonumber\\
&+\cos\left(\phi_1^{\2}+\phi_4^{\2}+\phi_1^{\4}+\phi_4^{\4}\right)\nonumber\\
&+\cos\left(\phi_2^{\1}-\phi_3^{\1}+\phi_2^{\2}-\phi_3^{\2}\right)\nonumber\\
&+\cos\left(\phi_2^{\1}-\phi_3^{\1}+\phi_2^{\3}-\phi_4^{\3}\right)\nonumber\\
&+\cos\left(\phi_2^{\3}-\phi_3^{\3}+\phi_2^{\4}-\phi_3^{\4}\right)\nonumber\\
&+\cos\left(\phi_2^{\2}-\phi_3^{\2}+\phi_2^{\4}-\phi_3^{\4}\right),
\label{type-1 Higgs}
\end{align}
where the subscript Arabic numerals label the different components of a specific quantum wire, and the superscript Roman numerals label the different quantum wires within each plaquette. 

A subtle point that needs additional care is that in the context of decoherence, all Higgs terms in doubled space should be able to be mapped to Kraus operators of some quantum channels in the physical Hilbert space. Our Higgs terms chosen here satisfy this requirement. Consider the first term in Eq. (\ref{type-1 Higgs}), it can be mapped to the following Kraus operators in the physical Hilbert space, 
\begin{align}
K_1=\cos(\varphi_1^\1+\varphi_1^\2),~K_2=\sin(\varphi_1^\1+\varphi_1^\2),
\end{align}
where $\varphi_1^{\1,\2}$ is mapped to $\phi_1^{\1,\2}$ and $\phi_4^{\1,\2}$ in the doubled Hilbert space by the Choi–Jamiołkowski isomorphism. Similarly, we can check all other terms in Eq. (\ref{type-1 Higgs}) can be mapped back to some Kraus operators in the physical Hilbert space. Therefore, we have obtained an explicitly wire construction for a type-\2 intrinsic SASPT in the doubled space.

\bibliography{references}

\end{document}